\documentclass[a4paper,12pt]{article}
\usepackage[dvips]{graphics}
\usepackage{epsfig}
\usepackage{psfrag}
\usepackage[psamsfonts]{amssymb}
\usepackage{amsmath}
\usepackage{indentfirst}
\usepackage{amssymb}
\usepackage{wrapfig}

\usepackage{graphicx}
\usepackage{amsmath}
\usepackage{amssymb}
\usepackage{theorem}
\usepackage{url}
\newtheorem{theorem}{Theorem}
\renewcommand{\thetheorem}{\arabic{section}.\arabic{theorem}}
\newtheorem{remark}{Remark}
\renewcommand{\thetheorem}{\arabic{section}.\arabic{theorem}}
\newtheorem{definition}{Definition}
\renewcommand{\thetheorem}{\arabic{section}.\arabic{theorem}}

\renewcommand{\thetheorem}{\arabic{section}.\arabic{theorem}}
\newtheorem{example}{Example}
\renewcommand{\thetheorem}{\arabic{section}.\arabic{theorem}}
\newtheorem{Assumption}{Assumption}
\renewcommand{\thetheorem}{\arabic{section}.\arabic{theorem}}

\begin{document}

\rule{0mm}{50mm}

\centerline{\Large Time-Bridge Estimators of Integrated Variance}

\rule{0mm}{20mm}

\centerline{\large A. Saichev$^{1,3}$, D. Sornette$^{1,2}$}

\rule{0mm}{20mm}

\centerline{$^{1}$\footnotesize ETH Zurich -- Department of Management, Technology and Economics, Switzerland}

\centerline{$^{2}$\footnotesize Swiss Finance Institute, 40, Boulevard du Pont-d' Arve, Case Postale~3, 1211 Geneva 4, Switzerland}

\centerline{$^{3}$\footnotesize Nizhni Novgorod State University -- Department of Mathematics, Russia.}

\rule{0mm}{60mm}

\centerline{E-mail addresses: saichev@hotmail.com \& dsornette@ethz.ch}

\clearpage

\rule{0mm}{30mm}

\centerline{\Large Time-Bridge Variance Estimators}

\rule{0mm}{10mm}

\begin{abstract}
We present a set of log-price integrated variance estimators, equal to
the sum of open-high-low-close bridge estimators of spot variances
within $n$ subsequent time-step intervals. The main characteristics of some
of the introduced estimators is to take into account the information
on the occurrence times of the high and low values. The use of the high's
and low's of the bridge associated with the original process makes the estimators
significantly more efficient that the standard realized variance estimators and its
generalizations. Adding the information
on the occurrence times of the high and low values improves further the
efficiency of the estimators, much above those of the well-known realized variance estimator
and those derived from the sum of Garman and Klass spot variance estimators.
The exact analytical results are derived for
the case where the underlying log-price process is an It\^o stochastic process.
Our results suggests more efficient ways to record financial prices
at intermediate frequencies.
\end{abstract}

\rule{0mm}{10mm}

Didier Sornette

Department of Management, Technology and Economics

(D-MTEC, KPL F38.2) ETH Zurich

Kreuzplatz 5

CH-8032 Zurich

Switzerland

\clearpage

\section{Introduction}

The integrated variance is a crucial risk indicator of the
stochastic log-price process within specific time intervals.
Most of the existing high-frequency
integrated variance estimators are modifications of
the well-known realized volatility (see, for instance,  Andersen et al. (2003), A\"it-Sahalia (2005), Zhang et al. (2005)), and are based on the knowledge of the open and close prices of
$n$ time-step intervals dividing the whole time interval of interest.
Another common practice to estimate the variance
of a log-price process is to use not two (open-close) log-prices within
a given time-step, but four values, the so-called the open-high-low-close (OHLC)
of the log-prices. Well-known examples are
the Garman and Klass (G\&K) (1980) and Parkinson (Park) (1980) spot variance estimators.

The main goal of this paper is to demonstrate the efficiency of bridge OHLC integrated variance estimators,
that use the knowledge of the high and low values of the bridge process
derived from the original log-price process, as well as possibly
the random occurrence times of these extrema within each time-step interval.
We compare the efficiencies of these time-OHLC bridge estimators with
the efficiency of the standard realized variance and with the efficiency of
the integrated variance estimators based on the G\&K estimators of
the variance within each elementary time-step interval.
We show that some time-OHLC integrated variance estimators achieve
a very significant improvement in efficiency compared with the
realized variance and the G\&K integrated variance estimators.
Another remarkable property of the proposed time-OHLC bridge estimators is that they
depend much less on the drift of the log-price process than the
realized variance and G\&K integrated variance estimators. This has the
great advantage of essentially removing the biases that affect the standard
estimators, given that the drift (expected return) is in general the most poorly
constrained statistical variable.
We compare the efficiencies of the introduced integrated variance estimators
using the It\^o process as our workhorse to model the stochastic behavior of log-prices.

Present databases record either all prices associated with transactions
or prune the data to keep the OHLC at given time steps, for instance,
seconds, minutes or days. The later records giving the OHLC of the realized log-prices
do not allow the reconstruction of the OHLC (and even less the occurrence times
of the high's and low's) for the associated bridge process in each elementary interval.
Of course, one could construct the OHLC and any other useful information
from the full time series of all transaction prices. But then, one could question
the value of deriving new estimators based on a reduced
information set. Therefore, the present paper can be considered as a
normative exercise to learn about the fundamental limits of integrated
variance estimators. Our results are also useful in suggesting
more efficient ways to record financial prices
at intermediate frequencies: instead of recording the OHLC at the daily scale for instance,
we propose that data centers and vendors should store to open and close of the real log-price
and the high and low of the corresponding bridge in each day
(or in any other chosen frequency). Our
calculations below show that this information, which has the same cost
and is as easy to obtain at the end of the day from the high frequency data,
provides much more efficient estimators of the variance that can be stored
for future use. The same conclusion holds true for other risk measures
beyond variance such as higher order moments, but this is not explored in the present paper.

The paper is organized as follows. Section~2 describes the properties of the
well-known realized variance estimator, which we need in order to compare its efficiency
with the efficiencies of the suggested time-OHLC bridge integrated variance estimators.
Section~3 is devoted to the discussion of the efficiencies of the simple bridge integrated variance estimators, illustrating the comparative efficiency and unbiasedness of the bridge integrated variance estimators. This section written in a pedagogical style gradually introduces the readers in the area
of homogeneous most efficient variance estimators. Section~4 provides a detailed analysis of
the efficiency of the OHL and time-OHLC bridge integrated variance estimators,
which turn out to be significantly more efficient than the realized variance and the G\&K integrated variance estimators. Section~5 describes the results of numerical simulations demonstrating
the comparative efficiency of the proposed estimators. Section~6 concludes.
The paper is completed by three appendix.  Appendix A presents the essential properties
of the canonical bridge. Appendix B derives the joint probability density function (pdf)
of the high value and of its occurrence time. Appendix C derives and gives the statistical
properties of the joint distribution of the high and low values and of the occurrence time
of the last extremum for the canonical bridge.

\section{Realized variance and beyond}

Henceforth, we assume that the log-price $X(t)$ of a given security follows an It\^o process
\begin{equation}\label{itoprdifeq}
d X(t) = \mu(t) dt +\sigma(t) d W(t) , \qquad X(0) = X_0 ,
\end{equation}
where $W(t)$ is a realization of the standard Wiener process,
while $\mu(t)$ is the drift process, and $\sigma^2(t)$ is the instantaneous
variance of the log-price process $X(t)$.

\subsection{Definitions and basic properties of realized variance}

Let us provide first some basic definitions and properties.
\begin{definition}
\textnormal{The integrated variance of the process $X(t)$ within the time interval $t\in(0,T)$ is
\begin{equation}\label{intvardef}
D(T):= \int_0^T \sigma^2(t) dt ~.
\end{equation}
}
\end{definition}

\begin{definition}
\label{htyju}
\textnormal{The spot variance is defined within the time-step interval
\begin{equation}\label{sampintdef}
\mathbb{S}_i : ~ (t_{i-1},t_i] ~,
\end{equation}
by
\begin{equation}\label{drvaridef}
\hat{D}_\text{real}\{X(t):t\in\mathbb{S}_i\}:= (X_i-X_{i-1})^2 , ~~ X_i:= X(t_i) ,~~  t_i := i \Delta ~,
~~ \Delta ={T\over n}~.
\end{equation}
}
\end{definition}

\begin{definition}
\textnormal{The well-known
statistical estimator of the integrated variance is the so-called realized variance defined as
\begin{equation}\label{realizevar}
[X,X]_T := \sum_{i=1}^n \hat{D}_\text{real}\{X(t):t\in\mathbb{S}_i\}~.
\end{equation}
}
\end{definition}

\begin{remark}
For It\^o processes \eqref{itoprdifeq} and for $n\to\infty$, it is well-known
that the realized variance converges in probability to the integrated one.
\end{remark}
However, for real data, the number $n$ of available data points  is always limited,
ultimately by the discreteness of the transaction flow and the associated
microstructure noise. Such structures, which are not taken into account
in the It\^o log-price model, can be neglected in the use of
the realized variance estimator if the discrete time step $\Delta$ is much
larger than the inverse of the  mean frequency $\nu$  of the tick-by-tick transactions,
so that $n \ll \nu T$.

\begin{Assumption}\label{supone}
\textnormal{While $\Delta \gg  1/\nu$, we assume that  $\Delta$
is sufficiently small in comparison with the time scales
over which the drift process $\mu(t)$ and the instantaneous variance $\sigma^2(t)$ vary,
so that one may replace the original It\^o process \eqref{itoprdifeq} by
Wiener processes with drift
\begin{equation}\label{itoprdifeqsimpl}
\begin{array}{c}
d X^i(t) \simeq \mu_i dt +\sigma_i d W(t) , \qquad X^i(t_{i-1}) = X_{i-1} , \qquad t\in\mathbb{S}_i ,
\\[3mm]
\mu_i = \text{const} , \qquad \sigma_i =\text{const}~ .
\end{array}
\end{equation}
}
\end{Assumption}

Consider the special case of the Wiener process with drift
\begin{equation}\label{xzerodef}
X(t,\mu,\sigma) = \mu t+\sigma W(t) ~.
\end{equation}
Using the scale-invariance property of the Wiener process, the
following identity holds in law (represented by the symbol $\sim$)
\begin{equation}\label{spotrealidentlaw}
\hat{D}_\text{real}\{X(t,\mu,\sigma):t\in\mathbb{S}_i\} \sim
\sigma^2 \Delta [\gamma+W(1)]^2 =\sigma^2 \Delta \cdot X^2(1; \gamma) ~,
\end{equation}
where
\begin{equation}\label{xtgamdef}
X(t; \gamma) = \gamma t + W(t) , \qquad \gamma = {\mu \over \sigma} \sqrt{\Delta} , \qquad t\in(0,1) ~,
\end{equation}
is the \emph{canonical Wiener process with drift}.
Applying the identity in law \eqref{spotrealidentlaw} to the realized variance
expression \eqref{realizevar}, \eqref{drvaridef}, we obtain
\begin{equation}
[X,X]_T \sim \Delta \sum_{i=1}^n \sigma_i^2(\gamma_i + W_i)^2 ~,
\end{equation}
where $\{W_i\}$ are iid Gaussian variables $\mathcal{N}(0,1)$.
Accordingly, the expected value of the realized variance is
\begin{equation}
\text{E}\left[[X,X]_T\right] = \Delta\sum_{i=1}^n \sigma_i^2 (1+\gamma_i^2) , \qquad \gamma_i = {\mu_i \over \sigma_i} \sqrt{\Delta} ~.
\end{equation}
This recovers the well-known fact that the realized variance is in general biased
for non-zero drift, and is non-biased only for zero-drift ($\mu(t)\equiv 0$).

\subsection{Beyond realized variance with new estimated variance estimators $\hat{D}_\text{est}(T)$}

The essential idea of the present work is that it is possible to
improve on the realized variance estimator of the
integrated variance estimator, for a fixed $n\ll \nu T$ of time-steps with
durations $\Delta$, by replacing it by
\begin{equation}\label{homintvarestdef}
\hat{D}_\text{est}(T) = \sum_{i=1}^n \hat{D}_\text{est}\{X(t):t\in\mathbb{S}_i\}~ ,
\end{equation}
where the functional $\hat{D}_\text{est}\{X(t):t\in\mathbb{S}_i\}$ is
an improved estimator of the spot variance given by definition
\ref{htyju}. The  subscript \texttt{est} is used
to refer to some particular estimator and the subscript \texttt{real}
means that this estimator reduces to the realized variance estimator.

\begin{definition}\label{unbiasdef}
\textnormal{The estimator $\hat{D}_\text{est}(T)$ defined by \eqref{homintvarestdef} is
said to be unbiased if, for all intervals $i=1, ... , n$,
\begin{equation}
\text{E}\left[\hat{D}_\text{est}\{X(t):t\in\mathbb{S}_i\}\right] = \Delta \cdot\sigma_i^2~,
\label{th6hw}
\end{equation}
which implies
\begin{equation}
 \text{E}\left[\hat{D}_\text{est}(T)\right] = \Delta \sum_{i=1}^n \sigma^2_i~ .
\end{equation}
When there exists at least one interval $j$, such that condition (\ref{th6hw})
does not hold, the estimator is considered biased.
}
\end{definition}

\subsection{Estimator efficiency}

Let $\hat{D}_\text{est}(T)$ be some unbiased variance estimator.
We propose to quantify its efficiency in terms of the coefficient of variation
\begin{equation}\label{vareffdef}
\rho[\hat{D}_\text{est}(T)] = {\sqrt{\text{Var}[\hat{D}_\text{est}(T)]}\over \text{E}[\hat{D}_\text{est}(T)]} .
\end{equation}

As an illustration, the coefficient of variation of the
realized variance for a Wiener process with zero drift ($\mu(t)\equiv0$) is equal to
\begin{equation}\label{rhorealthrysigmas}
\rho\left[[X,X]_T\right] = \displaystyle\sqrt{2\sum_{i=1}^n \sigma_i^4} \Bigg/ \displaystyle \sum_{i=1}^n \sigma_i^2 .
\end{equation}

We will need the following theorem:
\begin{theorem}\label{lowbound}
The lower bound of the function
\begin{equation}
f(\boldsymbol{s}) := \sqrt{\sum_{i=1}^n s_i^2} \Big/ \displaystyle \sum_{i=1}^n s_i , \qquad \boldsymbol{s} = \{s_1, s_2,\dots,s_n\} , \qquad \forall  s_i > 0
\end{equation}
is equal to
\begin{equation}
\rho(n):= \inf_{\forall s_i>0} f(\boldsymbol{s}) = {1\over \sqrt{n}} ~.
\end{equation}
And this lower bound is attained iff all $s_i$ are identical: $s_i\equiv s>0$.
\end{theorem}

\emph{\textbf{Proof.}}
Let $\{s_i\}$ be a realization of some random variable $S$ with probabilities
$\Pr\{S=s_i\} = {1\over n} , ~ i=1,\dots,n$.
Expected and mean square values of the random variable $S$ are equal to
\begin{equation}
\text{E}\left[S\right] = {1\over n} \sum_{i=1}^n s_i , \qquad
\text{E}\left[S^2\right] = {1\over n} \sum_{i=1}^n s^2_i~ .
\end{equation}
Since,  for any random variable $S$,
the inequality $\sqrt{\text{E}\left[S^2\right]} \geqslant \text{E}\left[S\right]$
holds, this implies $ f(s)\geqslant {1\over \sqrt{n}}$.
The inequality becomes an equality  iff all $s_i\equiv s$ for $\forall s>0$.
\hfill $\blacksquare$

\smallskip

Applying this theorem to the right-hand-side of expression \eqref{rhorealthrysigmas}
shows that $\rho\left[[X,X]_T\right]$ satisfies the inequality
\begin{equation}\label{lowboundrvar}
\rho\left[[X,X]_T\right] \geqslant \rho_\text{real}(n), \qquad \rho_\text{real}(n) = \sqrt{{2 \over n}} ~,
\end{equation}
where the lower bound $\rho_\text{real}(n)$ of the efficiency is attained
only if all $\{\sigma_i\}$ are identical.

Below, we will compare the efficiencies of different estimators
via the comparison of their lower bounds
\begin{equation}
\rho_\text{est}(n) = \inf_{\forall\sigma_i} \rho_\text{est}[\hat{D}(T)]~.
\end{equation}

\section{Realized bridge variance estimators}

\subsection{Basic definitions}

An important motivation for the introduction of a new class
of so-called ``realized bridge variance estimators'' is to obtain
much reduced biases compared that of the realized variance
\eqref{realizevar} observed for nonzero drifts $\mu(t)\not\equiv 0$.

\begin{definition}
\textnormal{The bridge $Y(t,\mathbb{S}_i)$ in discrete time steps
of the original process $X(t)$ is defined by
\begin{equation}\label{bridgegendef}
Y(t,\mathbb{S}_i) := X(t) - X_{i-1} -{t-t_{i-1}\over\Delta} \left(X_i-X_{i-1}\right) , \qquad t\in\mathbb{S}_i ,
\end{equation}
where $X_i:= X(t_i)$,  $t_i := i \Delta$ and $\Delta ={T\over n}$.
}
\end{definition}

As an example, let $X(t)$ be the Wiener process with drift $X(t,\mu,\sigma)$
defined by \eqref{xzerodef}. Using the transition and scale invariant properties of
the Wiener process leads to
\begin{equation}
Y(t,\mathbb{S}_i) \sim \sigma \sqrt{\Delta} \left(W(\zeta) - \zeta W(1)\right), \ \quad \zeta = {t-t_{i-1}\over \Delta} \in(0,1]~.
\end{equation}
This means that the bridge $Y(t,\mathbb{S}_i)$ \eqref{bridgegendef} is identical in law to
\begin{equation}\label{ytprimycan}
Y(t,\mathbb{S}_i) \sim \sigma \sqrt{\Delta} ~Y(\zeta) ,
\end{equation}
where
\begin{equation}\label{bridgewienerdef}
Y(t) := W(t) - t\cdot W(1) , \qquad t\in(0,1] ,
\end{equation}
is the \emph{canonical bridge} whose basic properties are given in Appendix~A.

\begin{remark}
\textnormal{The canonical bridge $Y(t)$
is completely independent of the drift $\mu$. This property is the fundamental
reason for the better performance of the variance bridge estimators compared
with the realized variance: the biases and efficiencies of
bridge variance estimators do not depend on the drift $\mu$.
}
\end{remark}

In the following, we explore the statistical properties of
the bridge variance estimators
\begin{equation}\label{homintvarbridgestdef}
\hat{D}_\text{est}(T) = \sum_{i=1}^n \hat{D}_\text{est}\{Y(t,\mathbb{S}_i):t\in\mathbb{S}_i\} ~,
\end{equation}
obtained from the general expression (\ref{homintvarestdef})
by replacing the initial process $\{X(t_i)\}$ by its corresponding bridge $\{Y(t_i,\mathbb{S}_i)\}$.

\begin{definition}\label{defhom}
\textnormal{The estimator \eqref{homintvarbridgestdef} is called \emph{homogeneous}
if, when applied to the Wiener processes with drift \eqref{itoprdifeqsimpl},
the following identity in law holds
\begin{equation}\label{varestidinlaw}
\hat{D}_\text{est}\{Y(t,\mathbb{S}_i):t\in\mathbb{S}_i\} \sim \sigma^2_i \Delta \cdot \hat{d}_\text{est} ~,
\end{equation}
where
\begin{equation}\label{canspotvarestdef}
\hat{d}_\text{est}:=\hat{D}_\text{est}\{Y(t):t\in(0,1]\}
\end{equation}
is the \emph{canonical estimator} of the spot variance depending on
the canonical bridge $Y(t)$ \eqref{bridgewienerdef}.
Obviously, the estimator \eqref{homintvarbridgestdef} is unbiased if and only if
$\text{E}\left[\hat{d}_\text{est}\right] = 1$.
}
\end{definition}

\begin{theorem}
Under Assumption \ref{supone}, the lower bound of the efficiency of the unbiased homogeneous integrated bridge variance estimator \eqref{varestidinlaw} is
\textnormal{
\begin{equation}\label{bridgelowbound}
\rho_\text{est}(n) = \sqrt{\text{Var}\left[\hat{d}_\text{est}\right]\big/ n} ~,
\end{equation}
}
where \textnormal{$\text{Var}\left[\hat{d}_\text{est}\right]$} is the variance of
the canonical spot variance estimator \textnormal{$\hat{d}_\text{est}$} \eqref{canspotvarestdef}.
\end{theorem}

\emph{\textbf{Proof.}}
Under Assumption~\ref{supone}, the unbiased homogeneous bridge variance estimator \eqref{homintvarbridgestdef} is identical in law to
\begin{equation}
\hat{D}_\text{est}(T) \sim \Delta \sum_{i=1}^n \sigma_i^2 \cdot \hat{d}_\text{est}^i ~,
\end{equation}
where $\{\hat{d}^i_\text{est}\}$ are iid random variables with mean value $\text{E}\left[\hat{d}_\text{est}\right]=1$ and variance $\text{Var}\left[\hat{d}_\text{est}\right]$. Accordingly, the expected value and variance of the unbiased bridge variance estimator \eqref{homintvarbridgestdef} are equal to
\begin{equation}
\text{E}\left[\hat{D}_\text{est}(T)\right] = \Delta \sum_{i=1}^n \sigma_i^2 , \qquad
\text{Var}\left[\hat{D}_\text{est}(T)\right] = \Delta^2 ~ \text{Var}\left[\hat{d}_\text{est}\right] \sum_{i=1}^n \sigma_i^4 .
\end{equation}
Substitute these relations into \eqref{vareffdef}, we obtain
\begin{equation}
\rho\left[\hat{D}_\text{est}(T)\right] = \displaystyle\sqrt{\text{Var}\left[\hat{d}_\text{est}\right]\sum_{i=1}^n \sigma_i^4} \Bigg/ \displaystyle \sum_{i=1}^n \sigma_i^2 .
\end{equation}
Using theorem \ref{lowbound}, this yields the result \eqref{bridgelowbound}. \hfill $\blacksquare$

\subsection{Simplest bridge variance estimator}

Our first example of an homogeneous bridge variance estimator is
\begin{equation}\label{bridgwquadthetaest}
\hat{D}_\text{simple}(T) = \sum_{i=1}^n \hat{D}_\text{simple}\{Y(t,\mathbb{S}_i):t\in\mathbb{S}_i\},
\end{equation}
where the estimator of the spot variance is given by
\begin{equation}\label{simplspotbrvar}
\hat{D}_\text{simple}\{Y(t,\mathbb{S}_i):t\in\mathbb{S}_i\} = A Y^2(t_i(\eta)) , ~~
t_i(\eta) = t_{i-1}+ \eta \cdot\Delta , ~~ \eta\in(0,1) ~,
\end{equation}
and $A$ is a normalizing factor. The estimator $\hat{D}_\text{simple}(T)$ is
homogeneous and,
if relations \eqref{itoprdifeqsimpl} are valid, then
\begin{equation}
Y^2(t_i(\eta), \mathbb{S}_i) \sim \sigma^2_i \Delta \cdot Y^2_i(\eta) ~,
\label{rhwhy3}
\end{equation}
where $\{Y_i(\eta)\}$ are iid random variables that
are identical in law to the canonical bridge \eqref{bridgewienerdef}. Substituting
relation (\ref{rhwhy3}) into \eqref{bridgwquadthetaest} leads to the identity in law
\begin{equation}
\hat{D}_\text{simple}(T) \sim A \Delta  \sum_{i=1}^n \sigma^2_i Y_i^2(\eta) ~.
\end{equation}
The fact that the canonical bridge $Y(\eta)$ is Gaussian
with mean value $\text{E}[Y^2(\eta)] = \eta (1-\eta)$ implies that
the estimator \eqref{bridgwquadthetaest} is unbiased in the sense of
definition \ref{unbiasdef} if $A = 1\big/ \eta(1-\eta)$.
Accordingly, the variance of the estimator \eqref{bridgwquadthetaest} is
equal to the variance of the realized variance obtained for zero drift ($\mu(t)\equiv0$):
\begin{equation}
\text{Var}[\hat{D}_\text{simple}(T)] = 2 \Delta^2 \sum_{i=1}^n \sigma_i^4~ .
\end{equation}
This result means that the lower bound of the efficiency of the
simplest bridge estimator \eqref{bridgwquadthetaest} is equal to the lower bound of
the efficiency of the realized variance estimator at zero drift:
\begin{equation}
\rho_\text{simple}(n) = \rho_\text{real}(n) = \sqrt{{2 \over n}}~ .
\label{trjejujw}
\end{equation}
The shortcoming of the estimator \eqref{bridgwquadthetaest} is that it is actually less efficient than
the realized variance at zero drift in a sense discussed below.

\subsection{Comparative efficiencies of realized variance estimators}

\begin{definition}
\textnormal{Let the estimator of the spot variance
\[
\hat{D}_\text{est}\{X(t):t\in\mathbb{S}_i\} \qquad \text{or} \qquad \hat{D}_\text{est}\{Y(t,\mathbb{S}_i):t\in\mathbb{S}_i\}
\]
depends on $\kappa_\text{est}$ values of the process $X(t)$ or $Y(t,\mathbb{S}_i)$ at
$\kappa_\text{est}$  time-step
within the time interval $t\in\mathbb{S}_i$. The corresponding estimators of
the realized volatility $\hat{D}_\text{est}(T)$ \eqref{homintvarestdef} or
\eqref{bridgwquadthetaest} are then using
a total number $n_\text{eff} = \kappa_\text{est} \cdot n$ of time-steps.
}
\end{definition}

\begin{example}
\textnormal{The realized variance corresponds to $\kappa_\text{real}=1$.
Indeed, the two values $\{X_{i-1},X_i\}$ are used to
estimate the spot realized variance \eqref{drvaridef},
and the first value is excluded from the semi-closed interval $\mathbb{S}_i$ \eqref{sampintdef}.}
\end{example}

\begin{example}
\textnormal{For the simplest bridge estimator \eqref{bridgwquadthetaest} with \eqref{simplspotbrvar},
$\kappa_\text{simple}=2$. Indeed, the estimator \eqref{simplspotbrvar} depends on
the bridge $Y(t_i(\eta), \mathbb{S}_i)$ for $ t_i(\eta)\in\mathbb{S}_i$ and
$Y(t_i(\eta), \mathbb{S}_i)$ \eqref{bridgegendef} is defined
by the open and close values $\{X_{i-1},X_i\}$ of the original stochastic process $X(t)$.
Excluding the open value, this yields $\kappa_\text{simple}=2$.
}
\end{example}

\begin{example}
\textnormal{Consider the Garman \& Klass (G\&K) variance estimator based
on open, high, low and close prices, used as the spot variance estimator in
expression \eqref{homintvarestdef}:
\begin{equation}\label{spvargk}
\begin{array}{c} \displaystyle
\hat{D}_\text{GK}\{X(t):t\in\mathbb{S}_i\} = k_1 (H_i-L_i)^2 - k_2 (C_i (H_i-L_i)-2H_i L_i) - k_3 C_i^2 ,
\\[3mm]\displaystyle
k_1 =0.511 , \qquad k_2 = 0.019 , \qquad k_3 = 0.383 ,
\end{array}
\end{equation}
where $\{O_i,C_i,H_i,L_i\}$ are the open, close, high and low values
\[
O_i = X_{i-1} , \quad C_i=X_i, \quad
H_i = \sup_{t\in\mathbb{S}_i} [X(t)-O_i] , \quad L_i = \inf_{t\in\mathbb{S}_i} [X(t)-O_i] .
\]
Excluding the open value leads to $\kappa_\text{GK}=3$.
}
\end{example}

\begin{definition}
\textnormal{We characterize the efficiencies of the novel
variance estimators by comparing them with that of the standard realized variance
estimator.  The corresponding comparative efficiency $\mathcal{R}_\text{est}$ is constructed
as the ratio of the lower bounds of the efficiencies of
the realized variance and novel variance estimator:
\begin{equation}
\mathcal{R}_\text{est} = {\rho_\text{real} (\kappa_\text{est}\cdot n) \over \rho_\text{est}(n)} ~.
\end{equation}
}
\end{definition}
Putting in this expression $\rho_\text{real}(n)$
given by equation \eqref{lowboundrvar} and $\rho_\text{est}(n)$ given by
expression \eqref{bridgelowbound} yields
\begin{equation}\label{ratiomcalrexpr}
\mathcal{R}_\text{est} = \sqrt{{2\over\kappa_\text{est}\cdot \text{Var}[\hat{d}_\text{est}]}}~ .
\end{equation}

\begin{remark}
\textnormal{For a given duration $T$ used to define the integrated variance \eqref{intvardef},
relation \eqref{ratiomcalrexpr} takes into account that
the typical waiting time between successive data samples is given by
$\Delta_\text{eff} \simeq T \big/ n_\text{eff}$. Such waiting time
should be approximately the same for the different generalized
variance estimators proposed below, leading
to similar distortions to the adequacy of the It\^o process \eqref{itoprdifeq}
in its ability to describe the real price process in the presence of
discrete tick-by-tick and other microstructure noise.
}
\end{remark}

\begin{example}
\textnormal{Let us come back to the simple variance estimator based
on expression (\ref{simplspotbrvar}) for
$\hat{D}_\text{simple}\{Y(t,\mathbb{S}_i):t\in\mathbb{S}_i\}$.
The result (\ref{trjejujw}) is equivalent to $\text{Var}[\hat{d}_\text{simple}]=2$.
Substituting this value in \eqref{ratiomcalrexpr} yields
\begin{equation}
\mathcal{R}_\text{simple} = {1\over\sqrt{\mathstrut\kappa_{\text{simple}}}}= {1\over\sqrt{2}} \simeq 0.707~ .
\end{equation}
The efficiency of the simplest bridge estimator is smaller than
that of the realized variance.
}
\end{example}

\begin{example}
\textnormal{Let us evaluate the comparative efficiency of the generalized realized variance
estimator based on the spot G\&K variance estimator in the case of zero drift $\mu(t)\equiv 0$.
It is known that the variance of the spot G\&K variance estimator
given by \eqref{spvargk} is equal to
\begin{equation}
\text{Var}\left[\hat{D}_\text{GK}\{X(t):t\in\mathbb{S}_i\}\right] = \sigma_i^2 \Delta \cdot 0.2693  \quad \Rightarrow \quad \text{Var}\left[\hat{d}_\text{GK}\right] =0.2693~.
\end{equation}
This gives
\begin{equation}\label{gkcompeff}
\mathcal{R}_\text{GK} = \sqrt{{2\over\mathstrut\kappa_\text{GK}\cdot 0.2693}} = \sqrt{{2\over\mathstrut 3\cdot 0.2693}} \simeq 1.573~.
\end{equation}
Therefore, for zero drift, the G\&K realized variance estimator is  approximately 1.6
times  more efficient than the realized variance estimator.
}
\end{example}

\subsection{High bridge variance estimator}

The fact that the G\&K realized variance estimator based on open-high-low-close
prices is significantly more efficient than the standard realized variance,
at least for It\^o process $X(t)$ \eqref{itoprdifeq} with zero drift $\mu(t)\equiv 0$,
suggests to study other estimators using different combinations of the
open-high-low-close prices. Let us start by analyzing the simplest
case of what we will refer to as the
``high bridge variance estimator'', defined through its spot variance given by
\begin{equation}\label{highspotestdef}
\hat{D}_\text{high}\{Y(t,\mathbb{S}_i):t\in\mathbb{S}_i\} = A\cdot H_i^2~ ,
\end{equation}
where $A$ is normalizing factor and
\begin{equation}
H_i = \sup_{t\in\mathbb{S}_i} Y(t,\mathbb{S}_i) ,
\end{equation}
is the high value of the bridge $Y(t,\mathbb{S}_i)$.
Note that we use here the same notation for the high value of
the bridge $Y(t,\mathbb{S}_i$) as for that of the original process $X(t)$,
hoping that this will not give rise to any confusion.

It follows from \eqref{ytprimycan} that
\begin{equation}
\hat{D}_\text{high}\{Y(t,\mathbb{S}_t):t\in\mathbb{S}_t\} \sim \sigma^2_i \Delta \cdot \hat{d}_\text{high} , \qquad \hat{d}_\text{high} =A H^2 ~,
\end{equation}
where the high value $H$ of the canonical bridge $Y(t)$ \eqref{bridgewienerdef}
has the following probability density function (pdf)
\begin{equation}
\varphi_\text{high}(h) = 4 h e^{-2 h^2} , \qquad h>0~ .
\label{juki5ikre}
\end{equation}
The derivation of the pdf (\ref{juki5ikre})
is given in Jeanblanc et al. (2009) (see also the derivations presented
in Appendix~B). Accordingly, the expected value and the variance of
the square of $H$ are given by
\begin{equation}
\text{E}\left[H^2\right] = {1\over2}, \qquad \text{Var}\left[H^2\right] = {1\over4} .
\end{equation}
In order for the high spot bridge variance estimator
to be unbiased, we have to choose in \eqref{highspotestdef}
the value $A=2$ for the normalizing factor. This gives
$\text{Var}\left[\hat{d}_\text{high} \right] = 1$.
With $\kappa_\text{high}=2$, we find that the comparative efficiency \eqref{ratiomcalrexpr}
of the high bridge realized variance estimator is
$\mathcal{R}_\text{high} = 1$.
Thus, the high bridge realized variance estimator
has the same efficiency as the standard realized variance.
But the advantage of the former is that, under Assumption \eqref{supone},
it is unbiased for any drift $\mu(t)\neq 0$.

\begin{remark}
\textnormal{Let us give the intuition for the above result, obtained despite the larger
value of $\kappa_\text{high}=2$ compared to $\kappa_\text{real}=1$.
The reason is that the pdf of the random variable $2H^2$ is narrower
than that of the random variable $W^2$ defining the spot realized variance at zero drift.
The same reason underlies the comparative efficiency of the G\&K as well the other
high and low bridge realized variance estimators discussed below.
The narrowness of the pdf's of high's and low's compared with the pdf's of
the increments of the original stochastic process $X(t)$ results from
a weak version of the Law of Large Numbers,
in the sense that the high's and low's incorporate significant additional information
about the underlying process within a given time-step, thus leading to narrower pdfs'.}
\end{remark}

\subsection{Time-high bridge variance estimator}

We now introduce a novel ingredient to improve further the estimation of the variance.
In addition to using only the high $H_i$ of the bridge $Y(t,\mathbb{S}_i)$,
we also assume that the time
$t_\text{high}^i$ of the occurrence of this high is recorded:
\begin{equation}
t_\text{high}^i : ~ H_i = Y(t_\text{high}^i,\mathbb{S}_i) ~.
\label{tjrukik5}
\end{equation}
The corresponding time-high bridge spot variance estimator is given by
\begin{equation}\label{highspotestdef}
\hat{D}_\text{est}\{Y(t,\mathbb{S}_i):t\in\mathbb{S}_i\}
= A\cdot s\left({t_\text{high}^i-t_{i-1}\over \Delta}\right) \cdot H_i^2~ ,
\end{equation}
where $A$ is a normalizing factor, while $s(t), t\in(0,1)$ is some function
that remains to be determined so as to
make the above spot variance estimator as  efficient as possible.
Before providing the solution of this problem,
let us note that the following identify in law follows from \eqref{ytprimycan}
\begin{equation}
\hat{D}_\text{est}\{Y(t,\mathbb{S}_i):t\in\mathbb{S}_i\} \sim \sigma^2_i \Delta \cdot \hat{d}_\text{est} ~,
\end{equation}
where
\begin{equation}\label{canthighest}
\hat{d}_\text{est} = A \cdot s(t_\text{high}) \cdot H^2
\end{equation}
is the canonical time-high bridge estimator of the spot variance,
$H$ is the high value of the canonical bridge $Y(t)$ \eqref{bridgewienerdef},
and $t_\text{high}$ is the corresponding time-point (\ref{tjrukik5}).

The expected value of the canonical estimator \eqref{canthighest} is equal to
\begin{equation}\label{alphlamtdef}
\text{E}\left[\hat{d}_\text{est}\right] =A \int_0^1 s(t) \alpha(t;2) dt, \quad \alpha(t;\lambda) := \int_0^\infty h^\lambda \varphi_\text{high}(h,t) d h
\end{equation}
where $\varphi_\text{high}(h,t)$ is the joint pdf of $H$ and $t_\text{high}$. Taking
\begin{equation}
A=1\Big/\int_0^1 s(t) \alpha(t;2) dt ~,
\end{equation}
we obtain an unbiased time-high canonical bridge estimator:
\begin{equation}
\hat{d}_\text{est} = { s(t_\text{high}) H^2\over \int_0^1 s(t) \alpha(t;2) dt} ~.
\label{r46jk7ki7}
\end{equation}
Its variance is
\begin{equation}\label{varunhatdthexpr}
\text{Var}\left[\hat{d}_\text{est}\right] = {\int_0^1 s^2(t)
\alpha(t;4) dt\over \left(\int_0^1 s(t) \alpha(t;2) dt\right)^2} - 1 .
\end{equation}

\begin{theorem}\label{thvarht}
The function $s(t)$ that minimizes the variance \eqref{varunhatdthexpr}
of the unbiased time-high canonical bridge estimator (\ref{r46jk7ki7}) is
\begin{equation}\label{ststarexpr}
s_\textnormal{t-high}(t) = {\alpha(t;2)\over \alpha(t;4)} .
\end{equation}
The corresponding minimal variance is equal to \textnormal{
\begin{equation}
\begin{array}{c}\displaystyle
\text{Var}\left[{ s_\text{t-high}(t_\text{high}) H^2\over \int_0^1 s(t) \alpha(t;2) dt}\right] =
\inf_{\forall \, s(t)} \text{Var}\left[\hat{d}_\text{est}\right] = {1\over\mathcal{E}_\text{t-high}} -1 ,
\\[5mm]\displaystyle
\mathcal{E}_\text{t-high} = \int_0^1 {\alpha^2(t;2)\over \alpha(t;4)} dt .
\end{array}
\end{equation}
}
\end{theorem}

\emph{\textbf{Proof.}} We use the Schwarz inequality
\begin{equation}
\left(\int_0^1A(t) B(t) dt\right)^2 \leqslant \int_0^1 A^2(t) dt \int_0^1 B^2(t) dt
\end{equation}
with
\begin{equation}
A(t) = s(t) \sqrt{\alpha(t;4)} , \qquad B(t) = {\alpha (t;2) \over \sqrt{\alpha(t;4)}} ~,
\end{equation}
to obtain
\begin{equation}
\left(\int_0^1 s(t) \alpha(t;2) dt\right)^2 \leqslant \int_0^1 s^2(t) \alpha(t;4)dt \int_0^1 {\alpha^2(t;2)\over\alpha(t;4)} dt .
\end{equation}
After simple transformations, we rewrite the last inequality in the form
\begin{equation}
\text{Var}\left[\hat{d}_\text{t-high}\right] = {\int_0^1 s^2(t) \alpha(t;4) dt\over \left(\int_0^1 s(t) \alpha(t;2) dt\right)^2} - 1 \geqslant {1\over  \int_0^1 {\alpha^2(t;2)\over\alpha(t;4)} dt} -1 ~.
\label{heyjujewrt}
\end{equation}
The equality in (\ref{heyjujewrt}) is reached by substituting in it $s(t)=s_\text{t-high}(t)$
given by expression \eqref{ststarexpr}. \hfill $\blacksquare$

\smallskip

The joint pdf of $H$ and $t_\text{high}$ is derived in Appendix~B and reads
\begin{equation}
\varphi_\text{high}(h,t) = \sqrt{2\over\pi} {h^2\over\sqrt{t^3 (1-t)^3}} \exp\left( - {h^2\over 2t(1-t)}\right) , \qquad h>0, \qquad t\in(0,1)~.
\end{equation}
Substituting this expression for $\varphi_\text{high}(h,t)$ into \eqref{alphlamtdef} yields
\begin{equation}
\alpha(t;\lambda) = {2\over\sqrt{\pi}} \left[2 t (1-t) \right]^{\lambda\over2} \Gamma\left({3+\lambda\over2}\right)~ .
\end{equation}
Therefore,
\begin{equation}
s_\text{t-high}(t) = {1\over 5 t (1-t)} , \qquad \mathcal{E}_\text{t-high} = {3\over5} \quad \Rightarrow \quad \text{Var}\left[\hat{d}_\text{t-high}\right] = {2\over3}~ ,
\end{equation}
and
\begin{equation}\label{mathrthigh}
\mathcal{R}_\text{t-high} = \sqrt{3\over2} \simeq 1.225 ~.
\end{equation}
Thus, the time-high bridge realized variance estimator is
less efficient than the corresponding G\&K estimator at zero drift, but is more efficient than
the realized variance.

\begin{remark}
\textnormal{The numerical result  \eqref{mathrthigh}
takes into account that the use of  $t^i_\text{high}$ does not
increase the number of sample values used in the spot estimator \eqref{highspotestdef}.
Thus,  $\kappa_\text{t-high} = \kappa_\text{high} = 2$.
}
\end{remark}

\section{Bridge time-high-low estimators}

\subsection{Bridge Parkinson estimator}

\begin{definition}
\textnormal{
The bridge realized variance estimator \eqref{homintvarbridgestdef}
that uses as spot variance estimator
\begin{equation}
\hat{D}_\text{bPark}\{Y(t,\mathbb{S}_i):t\in\mathbb{S}_i\} = A\cdot (H_i-L_i)^2
\label{thwrthynthkn}
\end{equation}
is called the \emph{bridge Parkinson estimator}. In expression (\ref{thwrthynthkn}),
$H_i$ and $L_i$ are the high and low values of the bridges $Y(t,\mathbb{S}_i)$ \eqref{bridgegendef}.
}
\end{definition}

The bridge Parkinson estimator is identical in law to
\begin{equation}\label{canparkest}
\hat{D}_\text{bPark}\{Y(t,\mathbb{S}_i):t\in\mathbb{S}_i\} \sim \sigma_i^2 \Delta \cdot \hat{d}_\text{bPark} , \qquad \hat{d}_\text{bPark} = A\cdot (H-L)^2 ,
\end{equation}
where $H$, $L$ are the high and low values of the canonical bridge $Y(t)$ \eqref{bridgewienerdef}.
The joint pdf of $H$ and $L$ have been derived by Saichev et al. (2009) and reads
\begin{equation}\label{mathrhldef}
\begin{array}{c}\displaystyle
\varphi(h,\ell) =
\sum_{m=-\infty}^\infty m
\left[ m \mathcal{I}(m(h-\ell)) + (1-m) \mathcal{I}(m(h-\ell)+\ell)\right] ,
\\[4mm]\displaystyle
\mathcal{I}(h) = 4 (4 h^2 - 1)~ e^{-2 h^2} .
\end{array}
\end{equation}

It will be clear below that it is convenient to describe
the joint statistical properties of the  high $H$ and low $L$ by using polar coordinates
\begin{equation}\label{polarcodef}
H = R \cos\Theta , \qquad L = R \sin\Theta , \qquad R\in(0,\infty) , \qquad \theta\in \left(-{\pi\over2},0\right)~ .
\end{equation}
Accordingly, we rewrite the canonical estimator \eqref{canparkest} in the form
\begin{equation}
\hat{d}_\text{bPark} = A R^2 \left(1-\sin2\Theta\right) ~.
\end{equation}
Choosing the constant $A$ that makes the estimator (\ref{canparkest}) unbiased, we obtain
\begin{equation}\label{alpparkdef}
\begin{array}{c}\displaystyle
\hat{d}_\text{bPark} = {R^2 (1-\sin2\Theta)\over \int_{-\pi/2}^0 (1-\sin2\theta) \alpha(\theta;2) d\theta} ,
\\[5mm]\displaystyle
\alpha(\theta;\lambda) = \int_0^\infty r^{\lambda+1}\varphi(r\cos\theta,r\sin\theta) dr .
\end{array}
\end{equation}
Substituting expression \eqref{mathrhldef} yields
\begin{equation}
\begin{array}{c}\displaystyle
\alpha(\theta;\lambda) =
\\ \displaystyle
\sum_{m=-\infty}^\infty m
\left[ m \beta(m(\cos\theta-\sin\theta);\lambda) + (1-m) \beta(m(\cos\theta-\sin\theta)+\sin\theta;\lambda)\right] ,
\\[3mm]\displaystyle
\beta(y;\lambda) = {C(\lambda)\over |y|^{2+\lambda}} , \qquad C(\lambda) = {1+\lambda\over \sqrt{2^\lambda}} \Gamma\left({2+\lambda\over 2}\right) .
\end{array}
\end{equation}
The variance of the canonical bridge Parkinson estimator is equal to
\begin{equation}
\text{Var}\left[\hat{d}_\text{bPark}\right] = {\int_{-\pi/2}^0 (1-\sin 2\theta)^2 \alpha(\theta;4) d\theta\over \left(\int_{-\pi/2}^0 (1-\sin 2\theta) \alpha(\theta;2) d\theta\right)^2} -1 \simeq 0.2000~ .
\end{equation}
Substituting this value into \eqref{ratiomcalrexpr} and taking into account that
$\kappa_\text{bPark}=3$ for the bridge Parkinson estimator, we obtain the comparative efficiency
\begin{equation}\label{varefbpark}
\text{Var}\left[\hat{d}_\text{bPark}\right] = 0.2000 , \quad \kappa_\text{bPark} = 3 , \qquad \Rightarrow \qquad \mathcal{R}_\text{bPark} \simeq 1.823~,
\end{equation}
which means that the bridge Parkinson estimator is significantly more efficient than
the G\&K estimator at zero drift.

\begin{remark}
\textnormal{
We stress that the canonical estimator $\hat{d}_\text{bPark}$ is significantly different
 from the well-known canonical Parkinson estimator (see Parkinson (1980))
\begin{equation}
\hat{d}_\text{Park} = {(H-L)^2 \over 4 \ln 2}~ ,
\end{equation}
where $H$ and $L$ are the high and low values of the
canonical Wiener process with drift $X(t,\gamma)$ \eqref{xtgamdef}.
In contrast with the bridge Parkinson estimator \eqref{alpparkdef} which is unbiased for any $\gamma$,
the standard Parkinson estimator is biased at nonzero drift. Moreover,
the variance of the standard Parkinson estimator at zero drift is
\begin{equation}
\text{Var}\left[\hat{d}_\text{Park}\right] \simeq 0.4073~,
\end{equation}
which is approximately twice the variance of the bridge Parkinson estimator \eqref{varefbpark}.
}
\end{remark}

\subsection{Non-quadratic homogeneous estimators}

Until now, we have
considered homogeneous (in the sense of definition \ref{defhom}) high-low estimators
that are quadratic functions of the high and low values.
We now consider the more general class of
homogeneous estimators, whose spot variance estimators have the form
\begin{equation}\label{nonqspestdef}
\hat{D}_\text{est}\{Y(t,\mathbb{S}_i):t\in\mathbb{S}_i\} = \mathcal{D}_\text{est}(H_i,L_i) ~,
\end{equation}
where $\mathcal{D}_\text{est}(h,\ell)$ is an arbitrary homogeneous function of second order.

\begin{example}
\textnormal{To illustrate the notion of non-quadratic homogeneous functions of second order,
consider the typical example
\begin{equation}
\mathcal{D}_\text{est}(H_i,L_i) = {(H_i-L_i)^3\over \sqrt{H_i^2+L_i^2}}~ ,
\end{equation}
which satisfies the scaling property
\begin{equation}\label{scalhomfundef}
\mathcal{D}_\text{est}(\delta\cdot H_i,\delta \cdot L_i) \equiv \delta^2 \cdot \mathcal{D}_\text{est}(H_i,L_i) \qquad \forall \delta > 0 ~.
\end{equation}
}
\end{example}

The following theorem states that the spot variance estimator \eqref{nonqspestdef} satisfies
the relations \eqref{varestidinlaw}, \eqref{canspotvarestdef}
of definition~\ref{defhom} for homogeneous estimators.

\begin{theorem}
The spot variance estimator \eqref{nonqspestdef} is homogeneous in the sense of definition~\ref{defhom}.
\end{theorem}

\emph{\textbf{Proof.}}
Let $H_i$ and $L_i$ be the high and low values of the bridge $Y(t,\mathbb{S}_i)$.
Due to relation \eqref{ytprimycan} and Assumption \ref{supone}, the following identity in law holds
\begin{equation}
\{H_i, L_i\} \sim \sigma_i\sqrt{\Delta} \cdot \{H, L\} ,
\end{equation}
where $\{H, L\}$ are the high and low values
of the canonical bridge $Y(t)$ \eqref{bridgewienerdef}. Substituting
this last relation into \eqref{nonqspestdef} yields
\begin{equation}
\mathcal{D}_\text{est}(H_i,L_i) \sim \mathcal{D}_\text{est}(\sigma_i \sqrt{\Delta}H,\sigma_i \sqrt{\Delta} L) ~.
\end{equation}
Using the homogeneity of the function $\mathcal{D}(h,\ell)$, we rewrite
the previous relation in the form
\begin{equation}
\mathcal{D}_\text{est}(H_i,L_i) \sim \sigma_i^2 \Delta_\text{est} \cdot \mathcal{D}(H,L)~ ,
\end{equation}
which is analogous to expression \eqref{varestidinlaw}, where
the canonical estimator of the spot variance is equal to
\begin{equation}
\hat{d}_\text{est} = \mathcal{D}_\text{est}(H,L) ~.
\end{equation}
$\blacksquare$

Using the polar coordinates \eqref{polarcodef}, the canonical estimator $\hat{d}_\text{est}$ reads
\begin{equation}\label{canhlestpolar}
\hat{d}_\text{est} = \mathcal{D}_\text{est}(R\cos\Theta,R\sin\Theta)~ .
\end{equation}
Using the homogeneity of the function $\mathcal{D}_\text{est}$, we obtain
\begin{equation}
\hat{d}_\text{est} = R^2 \cdot s(\theta) , \qquad s(\theta)= \mathcal{D}_\text{est}(\cos\theta,\sin\theta)~ .
\end{equation}
Its expected value is equal to
\begin{equation}
\text{E}\left[\hat{d}_\text{est}\right] = \int_{-\pi/2}^0  s(\theta) \alpha(\theta;2) d\theta~ ,
\end{equation}
where the function $\alpha(\theta,\lambda)$ is given by the equality \eqref{alpparkdef}.
Thus, the homogeneous non-quadratic canonical estimator reads
\begin{equation}\label{canunbiasest}
\hat{d}_\text{est} = {R^2 s(\Theta)\over \int_{-\pi/2}^0  s(\theta) \alpha(\theta;2) d\theta} \qquad \Rightarrow \qquad \text{E}[\hat{d}_\text{est}] = 1 .
\end{equation}
Accordingly, the variance of the unbiased estimator is equal to
\begin{equation}\label{varhomest}
\text{Var}\left[\hat{d}_\text{est}\right] = {\int_{-\pi/2}^0 s^2(\theta) \alpha(\theta;4) d\theta\over \left(\int_{-\pi/2}^0  s(\theta) \alpha(\theta;2) d\theta\right)^2} - 1.
\end{equation}

One can easily prove the result analogous to theorem~\ref{thvarht}
that the minimum value of the variance \eqref{varhomest} of the canonical estimator
\eqref{canunbiasest} with respect to all possible functions $s(\theta)$ is given by
\begin{equation}\label{mevarhomhl}
\text{Var}\left[\hat{d}_\text{me}\right] =\inf_{\forall s(\theta)} \text{Var}\left[\hat{d}_\text{est}\right] =
{1\over\mathcal{E}_\text{me}} -1 , \qquad \mathcal{E}_\text{me} = \int_{-\pi/2}^0 {\alpha^2(\theta;2)\over \alpha(\theta;4)} d\theta ,
\end{equation}
where $\hat{d}_\text{est}$ is an arbitrary homogeneous canonical estimator of the form \eqref{canunbiasest}, while $\hat{d}_\text{me}$ is the corresponding most efficient estimator given by
\begin{equation}
\hat{d}_\text{me} = {1\over\mathcal{E}_\text{me}} ~ R^2
s_\text{me}(\Theta) , \qquad s_\text{me}(\theta) = {\alpha(\theta;2)\over\alpha(\theta;4)} ~.
\end{equation}
Calculating the numerical value of the integral in expression \eqref{mevarhomhl} yields
\begin{equation}
\hat{d}_\text{me} = 0.1974 , \qquad \kappa_\text{me} = 3 , \qquad \Rightarrow \qquad \mathcal{R}_\text{me} \simeq 1.838 ~,
\end{equation}
which shows a high efficiency compared with the standard realized variance.

\subsection{Time-high-low homogeneous estimator}

Let us consider the unbiased homogeneous time-high-low canonical estimator
\begin{equation}\label{unbiasestgendef}
\hat{d}_\text{est} = {R^2 s(\Theta,t_\text{last}) \over  \int_0^1 dt \int_{-\pi/2}^0 d\theta ~ s(\theta,t) \alpha_\text{last}(\theta,t;2) } ,
\end{equation}
where $s(\theta,t)$ is an arbitrary function, $t_\text{last} = \sup\{t_L,t_H\}$ is
the larger of the two times at which occur the high and low values of the canonical bridge
and $\alpha_\text{last}(\theta,t;\lambda)$ is given by \eqref{alphlasttexpr}
in Appendix \ref{jruki}.

It is easy to prove the result analogous to theorem~\ref{thvarht}
that the most efficient estimator of the form \eqref{unbiasestgendef} is
\begin{equation}\label{bthtmint}
\hat{d}_\text{t-me} = {R^2\over\mathcal{E}_\text{t-me}} ~ {\alpha_\text{last}(\Theta,t_\text{last};2) \over \alpha_\text{last}(\Theta,t_\text{last};4)} , \quad
\mathcal{E}_\text{t-me} = \int_0^1 dt \int_{-\pi/2}^0 d\theta ~ {\alpha^2_\text{last}(\theta,t;2) \over \alpha_\text{last}(\theta,t;4)} ,
\end{equation}
and the variance of this estimator is equal to
\begin{equation}
\text{Var}\left[\hat{d}_\text{t-me}\right] = {1\over\mathcal{E}_\text{t-me}} -1~ .
\end{equation}
The numerical calculation of $\mathcal{E}_\text{t-me}$ gives
\begin{equation}\label{varmathrtme}
\text{Var}\left[\hat{d}_\text{t-me}\right] \simeq 0.1873, \quad \kappa_\text{t-me} = 3 , \qquad \Rightarrow \qquad \mathcal{R}_\text{t-me} \simeq 1.887 .
\end{equation}
The estimator of the realized variance based on the canonical estimator \eqref{bthtmint} is significantly more efficient than that based on the G\&K estimator at zero drift.

\subsection{High-low-close bridge estimator}

Until now, we have not used explicitly the information contained in the close values $X_i$ \eqref{drvaridef}
of the time-step intervals $\mathbb{S}_i$ \eqref{sampintdef}.
The close values $X_i$ have been used only
for the construction of the bridge $Y(t,\mathbb{S}_i)$ \eqref{bridgegendef}.  It seems plausible
that taking into account explicitly the close values $X_i$ in the construction of spot variance estimators
may produce bridge realized variance estimators
$\hat{D}_\text{est}(T) = \sum_{i=1}^n \hat{D}_\text{est}\{Y(t,\mathbb{S}_i):t\in\mathbb{S}_i;X_i\}$
that are even more efficient than those considered until now.
We show that this is indeed the case by studying the example associated with
the spot variance estimator given by
\begin{equation}
\hat{D}_\text{est}\{Y(t,\mathbb{S}_i):t\in\mathbb{S}_i;X_i\} = \mathcal{D}_\text{est}(H_i,L_i,X_i)~ ,
\end{equation}
where $\mathcal{D}_\text{est}(h,\ell,x)$ is an arbitrary homogeneous function satisfying relation \eqref{scalhomfundef}. Due to its homogeneity, the following identity in law holds true
\begin{equation}\label{canhlxestdef}
\mathcal{D}_\text{est}(H_i,L_i,X_i) \sim \sigma_i^2 \Delta\cdot  \hat{d}_\text{est} , \qquad \hat{d}_\text{est}=\mathcal{D}_\text{est}(H,L,X) ,
\end{equation}
where $H$ and $L$ are the high and low values of the canonical bridge \eqref{bridgewienerdef},
while $X=\gamma+W$ is the close value of the underlying canonical Wiener process with drift \eqref{xtgamdef}. It is known (see, for instance, Jeanblanc et al. (2009)) that
the canonical bridge $Y(t)$ and $W$ are statistically independent.
Thus, the joint pdf $\varphi(h,\ell,x)$ of the three random variables $\{H,L,X\}$ is equal to
\begin{equation}
\varphi(h,\ell,x;\gamma) = {1\over\sqrt{2\pi}} \exp\left(-{(x-\gamma)^2\over2} \right) \varphi(h,\ell)~ ,
\end{equation}
where the joint pdf $\varphi(h,\ell)$ of high and low values is given by expression \eqref{mathrhldef}.

Analogously to \eqref{canhlestpolar}, it is convenient to represent
the canonical estimator $\hat{d}_\text{est}$ \eqref{canhlxestdef} in the spherical coordinate system
\begin{equation}
\begin{array}{c}
H= R\cos\Upsilon\cos\Theta , \qquad L= R\cos\Upsilon\sin\Theta , \qquad X = R \sin\Upsilon ,
\\[2mm] \displaystyle
\Upsilon \in (-\pi/ 2,\pi/ 2) , \qquad \Theta\in (-\pi/2,0) .
\end{array}
\end{equation}
Th canonical estimator $\hat{d}_\text{est}$ \eqref{canhlxestdef}
then takes the form
\begin{equation}
\hat{d}_\text{est} = R^2 s(\Theta, \Upsilon) ,
\end{equation}
where
\begin{equation}
s(\Theta, \Upsilon) = \mathcal{D}_\text{est}(\cos\Upsilon\cos\Theta,\cos\Upsilon\sin\Theta, \sin\Upsilon)~.
\end{equation}

Analogously to \eqref{unbiasestgendef} and \eqref{bthtmint},
the unbiased most efficient high-low-close canonical estimator is given by
\begin{equation}\label{canunbiasestme}
\hat{d}_\text{me-x} = {1\over\mathcal{E}_\text{me-x}} ~ R^2 s_\text{me-x}(\Theta,\Upsilon;\gamma) , \qquad s_\text{me-x}(\theta,\upsilon;\gamma) = {\alpha(\theta,\upsilon;2; \gamma)\over\alpha(\theta,\upsilon;4; \gamma)}~.
\end{equation}
The function $\alpha(\theta,\upsilon;\lambda; \gamma)$ is defined by the equality
 \begin{equation}
\alpha(\theta,\upsilon;\lambda; \gamma) = \int_0^\infty r^{\lambda+2} \varphi(r\cos\upsilon\cos\theta,r\cos\upsilon\sin\theta,r\sin\upsilon; \gamma) dr~ .
\end{equation}
The variance of the most efficient canonical estimator $\hat{d}_\text{me-x}$ is equal to
\begin{equation}
\text{Var}\left[\hat{d}_\text{me-x}\right] = {1\over\mathcal{E}_\text{me-x}} -1 ,
\end{equation}
with
\begin{equation}\label{mathemexdef}
\mathcal{E}_\text{me-x} = \int_{-\pi/2}^0d\theta \int_{-\pi/2}^{\pi/2}d\upsilon \cos\upsilon {\alpha^2(\theta,\upsilon;2; \gamma)\over\alpha(\theta,\upsilon;4; \gamma)} ~.
\end{equation}
The calculation of the integral \eqref{mathemexdef} for $\gamma=0$ gives
\begin{equation}
\text{Var}\left[\hat{d}_\text{me-x}\right] \simeq 0.1794 , \quad \kappa_\text{me-x} =3,  \qquad \Rightarrow \qquad \mathcal{R}_\text{me-x} \simeq 1.928~ .
\label{herhbh3t}
\end{equation}
This estimator is definitely better than the most efficient time-high-low canonical estimator,
as can be seen by comparing (\ref{herhbh3t}) with \eqref{varmathrtme}.

\subsection{Time-high-low-close bridge estimator}

The last example we present here is the realized variance estimator
that uses in each interval $\mathbb{S}_i$
the high and low values $H_i$, $L_i$
of the bridge $Y(t,\mathbb{S}_i)$ \eqref{bridgegendef},
the close value $X_i$ of the original stochastic process $X(t)$ and
the time instant $t_\text{last}^i= \sup\{t_L^i,t_H^i\}$ defined as
the larger of the two times at which occur the high and low values of the canonical bridge.

One can rigorously prove that, analogously to \eqref{canunbiasestme},
the homogeneous time-OHLC bridge canonical estimator that is most efficient
for some given value of $\gamma$ value is equal to
\begin{equation}\label{bthtmin}
\begin{array}{c}\displaystyle
\hat{d}_\text{t-me-x}(\Theta,\Upsilon,t_\text{last};\gamma) = R^2 s_\text{t-me-x}(\Theta,\Upsilon,t_\text{last};\gamma) ,
\\[4mm]\displaystyle
s_\text{t-me-x}(\theta,\upsilon,t;\gamma) = {1\over\mathcal{E}_\text{t-me-x}(\gamma)} ~ {\alpha(\theta,\upsilon,t;2; \gamma) \over \alpha(\theta,\upsilon,t;4; \gamma)} ,
\end{array}
\end{equation}
where
\begin{equation}\label{mathedef}
\mathcal{E}_\text{t-me-x}(\gamma) = \int_0^1 dt \int_{-\pi/2}^0 d\theta \int_{-\pi/2}^{\pi/2} d\upsilon~\cos\upsilon {\alpha^2(\theta,\upsilon,t;2; \gamma) \over \alpha(\theta,\upsilon,t;4; \gamma)}
\end{equation}
and
\begin{equation}\label{athuptlanonzeroga}
\alpha(\theta,\upsilon,t;\lambda; \gamma) = \int_0^\infty r^{\lambda+2} \varphi_\text{last}(r\cos\upsilon\cos\theta,r\cos\upsilon\sin\theta,r\sin\upsilon,t; \gamma) dr .
\end{equation}
The joint pdf $\varphi(h,\ell,x,t;\gamma)$ is
\begin{equation}\label{philasthelxtgam}
\varphi_\text{last}(h,\ell,x,t;\gamma) = {1\over\sqrt{2\pi}} \exp\left(-{(x-\gamma)^2\over2} \right) \varphi_\text{last}(h,\ell,t) ,
\end{equation}
where $\varphi_\text{last}(h,\ell,t)$ is given by expression \eqref{fhelltexpr}
in Appendix \ref{hyjywtbbr}.

\begin{remark}
\textnormal{Recall that
the parameter factor $\gamma$ \eqref{xtgamdef} is unknown, because both
the drift $\mu_i$ and the instantaneous variances $\sigma^2_i$ in equations \eqref{itoprdifeqsimpl}
are generally unknown. Therefore,
our strategy below is to choose, for definiteness, $\gamma=0$ and then explore
the dependence on $\gamma$ of the bias and efficiency of
the different  ``zero drift'' estimators. Accordingly, we will use below
the following shorthand notations, omitting the argument $\gamma$, such as
\[
\hat{d}_\text{t-me-x}(\Theta,\Upsilon,t):= \hat{d}_\text{t-me-x}(\Theta,\Upsilon,t;\gamma=0) .
\]
}
\end{remark}

The calculation of the integral \eqref{mathedef}, where $\alpha(\theta,\upsilon,t;\lambda)$
is given by expression \eqref{athuptlaga} in Appendix \ref{hruyjkikik}
yields for $\gamma=0$
\begin{equation}
\text{Var}\left[\hat{d}_\text{t-me-x}\right] = {1\over\mathcal{E}_\text{t-me-x}} -1 \simeq 0.1710, \quad \kappa_\text{t-me-x} =3, \quad \Rightarrow \quad \mathcal{R}_\text{t-me-x} \simeq 1.975~.
\end{equation}
This estimator is more efficient than all the previous one discussed until now.

\section{Numerical simulations and comments}

\subsection{Description of numerical simulations}

The goal of this section is to check by numerical simulations some
analytical results obtained above. Realizations of the canonical Wiener process $X(t;\gamma)$ \eqref{xtgamdef}
with drift for time $t\in [0,1]$ are obtained numerically as
cumulative sums of a number $I(t)=10^5$ of Gaussian summands, corresponding
to a discrete time step $\Delta = 10^{-5}$.
For each numerical realization, we calculate the values of the open-close spot variance canonical estimator, equal in this case to
\begin{equation}\label{realspotnumsimest}
\hat{d}_\text{real} = (\gamma+ W)^2~,
\end{equation}
and the values of the G\&K canonical estimator
\begin{equation}\label{gkspotnumsimest}
\hat{d}_\text{GK} = k_1 (H-L)^2 - k_2 (W (H-L)-2H L) - k_3 (\gamma+ W)^2~ ,
\end{equation}
where $H$ and $L$ are the high and low values of the simulated process $X(t;\gamma)$.

We also constructed numerical realizations of
the bridge process $Y(t)$ \eqref{bridgewienerdef} and calculated
the corresponding values of the canonical estimator $\hat{d}_\text{t-me-x}$ \eqref{bthtmin}.
This estimator depends on the function $\alpha(\theta,\upsilon,t;\lambda)$
defined by expression \eqref{athuptlaga} in Appendix  \ref{hruyjkikik}, which
is explicitly obtained by summing a double-infinite series (\ref{alphlasttexprprim}).
In practice, we estimate this double-sum by keeping only the $101$ first
terms in each dimension, corresponding to estimating $101 \times 101 \simeq 10^4$
summands in \eqref{athuptlaga}.

\begin{remark}
\textnormal{At first glance, it would seem that the
calculation of the G\&K estimator \eqref{gkspotnumsimest}, which needs only
a few simple arithmetic operations, is much easier than
the evaluation of the large number  of summands in the series \eqref{athuptlaga}
that define the estimator $\hat{d}_\text{t-me-x}$ \eqref{bthtmin}.
In our computerized world, it turns out that there is actually no significant difference from
the computational point of view.
}
\end{remark}

\subsection{Statistics of the estimators in the case of zero drift ($\gamma=0$)}

Figure~1 shows 5000 realizations of the open-close estimator $\hat{d}_\text{real}$ \eqref{realspotnumsimest},
of the G\&K estimator \eqref{gkspotnumsimest} and of the estimator $\hat{d}_\text{t-me-x}$ in the case
the Wiener process with zero drift ($\gamma=0$). It is clear that the last estimator is
the most efficient in comparison with the open-close and the G\&K estimators.
The expected values and variances of these three estimators
obtained by statistical averaging over $10^4$ samples are
\[
\begin{array}{c}
\text{E}[\hat{d}_\text{real}]\simeq 1.0110 , \qquad \text{E}[\hat{d}_\text{GK}] \simeq 1.0058, \qquad \text{E}[\hat{d}_\text{t-me-x}] \simeq 1.0001 ,
\\[1mm] \displaystyle
\text{Var}[\hat{d}_\text{real}]\simeq 1.9947 , \qquad \text{Var}[\hat{d}_\text{GK}] \simeq 0.2669, \qquad \text{Var}[\hat{d}_\text{t-me-x}] \simeq 0.1696 .
\end{array}
\]
These values are consistent with the theoretical analytical predictions
obtained in previous sections:
\[
\begin{array}{c}
\text{E}[\hat{d}_\text{real}]=\text{E}[\hat{d}_\text{GK}]= \text{E}[\hat{d}_\text{t-me-x}] = 1 ,
\\[1mm] \displaystyle
\text{Var}[\hat{d}_\text{real}] = 2 , \qquad \text{Var}[\hat{d}_\text{GK}] \simeq 0.2693, \qquad \text{Var}[\hat{d}_\text{t-me-x}] \simeq 0.1710 .
\end{array}
\]

In order to have truly comparable efficiencies of these realized variance estimators, bearing in mind that their effective sample sizes are different ($\kappa_\text{real}=1$, $\kappa_\text{GK}= \kappa_\text{t-me-x}=3$),
we performed moving averages with $r=30$ subsequent samples for the open-close estimator \eqref{realspotnumsimest} and with $r=10$ subsequent samples for the G\&K estimator \eqref{gkspotnumsimest} and estimator $\hat{d}_\text{t-me-x}$ \eqref{bthtmin}. Figure~2 presents there moving averages, which mimick
the normalized estimators of the integrated variance in the case where all instantaneous variances are the same ($\sigma_i^2=\sigma^2=\text{const}$). It is clear that the open-close estimator of the realized variance remains significantly less efficient than the G\&K estimator, and much less efficient than the most efficient estimator $\hat{d}_\text{t-me-x}$.

\subsection{$\gamma$-dependence of biases and efficiencies of canonical estimators}

In the previous subsection, we presented detailed calculations
of the comparative efficiency of unbiased variance estimators for
the particular case of Wiener processes with zero drift.
In real financial markets, the drift process $\mu(t)$ is
unknown and there is not reason for it to vanish. Thus, it is important to explore quantitatively
the dependence on the parameter $\gamma$ \eqref{xtgamdef} of
the biases and efficiencies of the spot variance canonical estimators
described above.

 We begin with the open-close spot variance canonical estimator $\hat{d}_\text{real}$ \eqref{realspotnumsimest}. It is easy to show that its expected value and variance
 are quadratic functions of $\gamma$:
\begin{equation}\label{opclexpvar}
\text{E}\left[\hat{d}_\text{real}\right] = 1+ \gamma^2 , \qquad \text{Var}\left[\hat{d}_\text{real}\right] = 2 + 4 \gamma^2 .
\end{equation}
The spot variance homogeneous time-open-high-low canonical bridge estimators,
such as the Park estimator $\hat{d}_\text{bPark}$ \eqref{alpparkdef} and
the time-high-low estimator $\hat{d}_\text{t-me}$ \eqref{bthtmint}, are unbiased for all $\gamma$:
\[
\text{E}\left[\hat{d}_\text{bPark}\right] = \text{E}\left[\hat{d}_\text{t-me}\right] \equiv 1~.
\]
Their variances do not depend on $\gamma$ at all:
\begin{equation}\label{parktmevars}
\text{Var}\left[\hat{d}_\text{bPark}\right] \simeq 0.2000 , \qquad \text{Var}\left[\hat{d}_\text{t-me}\right] \simeq 0.1873 \qquad \forall ~ \gamma.
\end{equation}

To obtain the $\gamma$-dependence of the biases and variances of
the G\&K canonical estimator $\hat{d}_\text{GK}$ \eqref{gkspotnumsimest} and
of the canonical estimator $\hat{d}_\text{t-me-x}$ \eqref{bthtmin},
we generate $10^4$ numerical realizations of the canonical Wiener process
$X(t,\gamma)$ \eqref{xtgamdef} with drift, for  $\gamma = 0; 0.1; \dots 1.5; 1.6$.
Then, we calculated the statistical averages and variances of
the corresponding $10^4$ realizations of the canonical estimators
$\hat{d}_\text{GK}$ and $\hat{d}_\text{t-me-x}$, which are shown in
figure~3. The continuous lines are respectively
the expected value \eqref{opclexpvar} of the open-close estimator $\hat{d}_\text{real}$
given by expression \eqref{realspotnumsimest} and the fitted curves
\[
\text{E}[\hat{d}_\text{est}] = a_\text{est} \gamma^2 + b_\text{est}
\]
for the averaged values of the canonical estimators
$\hat{d}_\text{GK}$ and  $\hat{d}_\text{t-me-x}$. Their fitted parameters are
\[
a_\text{GK} \simeq 0.126 , \qquad a_\text{t-me-x} \simeq 0.082 , \qquad
b_\text{GK} \simeq b_\text{t-me-x}\simeq 1 .
\]

Figure~4 shows the statistical average of
the variances of the canonical estimators $\hat{d}_\text{GK}$ and $\hat{d}_\text{t-me-x}$.
The two horizontal lines indicate the variance values \eqref{parktmevars}.
The continuous lines show the fitted curves
\[
\text{Var}[\hat{d}_\text{est}] = c_\text{est} \gamma^2 + d_\text{est}
\]
of the variances of the canonical estimators $\hat{d}_\text{GK}$ and  $\hat{d}_\text{t-me-x}$.
Their parameters are
\[
c_\text{GK} \simeq 0.089 , \qquad c_\text{t-me-x} \simeq 0.0272 , \qquad
d_\text{GK} \simeq 0.271 , \qquad b_\text{t-me-x}\simeq 0.170 .
\]

\subsection{Construction of general variance estimators}

We have introduced the canonical estimator $\hat{d}_\text{t-me-x}$
given by expression \eqref{bthtmin} that includes the information
on the value of the time  $t_\text{last} = \sup\{t_L,t_H\}$ defined as
the larger of the two times at which occur the high and low values of the canonical bridge.
It seems that the canonical estimator
\begin{equation}\label{ttmexcanest}
\begin{array}{c}\displaystyle
\hat{d}_\text{tt-me-x}(\Theta,\Upsilon,t_\text{high},t_\text{low};\gamma) = R^2 s_\text{tt-me-x}(\Theta,\Upsilon,t_\text{high},t_\text{low};\gamma) ,
\\[4mm]\displaystyle
s_\text{tt-me-x}(\theta,\upsilon,t_1,t_2;\gamma) = {1\over\mathcal{E}_\text{tt-me-x}(\gamma)} ~ {\alpha(\theta,\upsilon,t_1,t_2;2; \gamma) \over \alpha(\theta,\upsilon,t_1,t_2;4; \gamma)} ,
\end{array}
\end{equation}
taking into account both high's and low's and their corresponding occurrence
times ($t_\text{high}: ~ H = Y(t_\text{high}) , t_\text{low}: ~ L = Y(t_\text{low})$)
is even more efficient than the estimator \eqref{bthtmin}. In
expression \eqref{ttmexcanest}, we have used the notation
\begin{equation}
\alpha(\theta,\upsilon,t_1,t_2;\lambda, \gamma) = \int_0^\infty r^{\lambda+2} \varphi(r\cos\upsilon\cos\theta, r\cos\upsilon\sin\theta,r\sin\upsilon,t_1,t_2;\gamma) dr~ ,
\label{hyujuj}
\end{equation}
where $\varphi(h,\ell,x,t_1,t_2;\gamma)$ is the joint pdf of the
high-low-close-$t_\text{hight}$-$t_\text{low}$ random variables.

We have not explored the statistical properties of
the estimator \eqref{ttmexcanest} because we have made not yet the effort
of deriving the exact analytical expression of $\varphi(h,\ell,x,t_1,t_2;\gamma)$.
We can however construct the function $\alpha$ (\ref{hyujuj})
using statistical averaging:
\begin{equation}\label{alphanumsim}
\alpha(\theta,\upsilon,t_1,t_2;\lambda, \gamma) \cos\upsilon d\upsilon d\theta dt_1 dt_2 \simeq
{1\over K} \sum_{k=1}^K R^\lambda_k \boldsymbol{I} \left(\Upsilon_k, \Theta_k, t_{\text{high},k}, t_{\text{low},k}\right) ~.
\end{equation}
In this expression, the values $\{\Upsilon_k, \Theta_k, t_{\text{high},k}, t_{\text{low},k}\}$
are parameters of numerically simulated $k$-th sample of
the canonical Wiener process with drift $X(t,\gamma)$ \eqref{xtgamdef},
and $\boldsymbol{I}$ is the indicator of the set
\[
(\upsilon,\upsilon+d\upsilon)\times (\theta,\theta+d\theta)\times (t_1,t_1+dt_1)\times (t_2,t_2+dt_2)~.
\]

We would like to point out that it is possible to
construct the function $\alpha$ by an analogous statistical treatment
for more general log-price process that extend the Wiener process with drift
to include more adequately the micro-stricture noise,
the presence of heavy tails of returns and other stylized facts that
can be found for various financial assets.
In others words, relations such as \eqref{alphanumsim} offer the possibility
of constructing novel most efficient variance estimators of the form \eqref{ttmexcanest},
extending the standard approach of
econometricians looking for new constructions of efficient volatility estimators.
The requisite is to be able to simulate numerically the underlying stochastic process
that is representing a given financial asset dynamics. Then, the use of
statistical averaging, similar to \eqref{alphanumsim}, will
enable the construction of  high-frequency realized estimators
that use the most efficient estimators described above as elementary ``bricks''.

\section{Conclusion}

We have introduced a variety of integrated variance estimators,
based on the open-high-low values of the bridges $Y(t,\mathbb{S}_i)$ \eqref{bridgegendef}, and close values $X_i$ \eqref{drvaridef} of the underlying log-price process $X(t)$. The main peculiarity of some of the
introduced estimators is to take into account not only the high and low values but additionally
their occurrence time. This last piece of information lead to estimators that
are even more efficient. We discussed quantitatively the statistical properties of
the estimators for the class off It\^o model for the log-price stochastic process.

Our work opens the road to the construction of novel types
of integrated variance estimators of log-price processes of real financial markets
that take into account the microstructure noise, heavy power tails of returns, and chaotic jumps.

\vskip 0.3cm
{\bf Acknowledgements}: We are grateful to Fulvio Corsi for valuable discussions of some aspects of this paper.

\appendix
\setcounter{section}{0}
\setcounter{equation}{0}
\setcounter{theorem}{0}
\renewcommand{\theequation}{\thesection.{\arabic{equation}}}
\renewcommand{\thetheorem}{\thesection.{\arabic{theorem}}}
\renewcommand{\thesection}{\Alph{section}}

\section{Basic properties of the canonical bridge}

\setcounter{equation}{0}
\setcounter{theorem}{0}
\setcounter{remark}{0}
\setcounter{lemma}{0}
\setcounter{definition}{0}
\setcounter{example}{0}
\setcounter{Assumption}{0}

\subsection{Symmetry properties}

The canonical bridge $Y(t)$ \eqref{bridgewienerdef} exhibits the following
 time reversibility and reflection properties
\begin{equation}\label{reversdef}
Y(t) \sim Y(1-t) , \qquad Y(t) \sim -Y(t) .
\end{equation}
Some statistical consequences of these symmetry properties are as follows. Let
\begin{equation}\label{highlowdef}
H =\sup_{t\in(0,1)} Y(t) , \qquad L =\inf_{t\in(0,1)} Y(t) ,
\end{equation}
be the high and low values of the canonical bridge, while $t_\text{high}$ and $t_\text{low}$
are their corresponding occurrence times:
\begin{equation}\label{instmaxmin}
t_\text{high}: ~ H = Y(t_\text{high}) , \qquad t_\text{low}: ~ L = Y(t_\text{low}) .
\end{equation}

Consider the cumulative distribution (cdf)
\[
\Phi_\text{high}(t) = \Pr\{t_\text{high}<t\}
\]
of the occurrence time $t_\text{high}$ of the high value of
the canonical bridge. Due to the reversibility property \eqref{reversdef},  one has
\begin{equation}\label{cdfsreverse}
\Pr\{t_\text{high}<t\} = \Pr\{t_\text{high}>1-t\} \quad \Rightarrow \quad \Phi_\text{high}(t) + \Phi_\text{high}(1-t) =1 .
\end{equation}
Accordingly, the pdf of $t_\text{high}$
\[
\varphi_\text{high}(t) := {d \Phi_\text{high}(t)\over dt}
\]
presents the symmetry
\begin{equation}\label{pdftsym}
\varphi_\text{high}(t) = \varphi_\text{high}(1-t) .
\end{equation}

Due to the reversibility property of the canonical bridge,
the cdf $\Phi_\text{low}(t)$ of $t_\text{low}$ \eqref{instmaxmin} coincides with
the cdf of $t_\text{high}$:
\[
\Phi_\text{low}(t) = \Phi_\text{high}(t) \quad \Rightarrow \quad \varphi_\text{low}(t) = \varphi_\text{high}(t) = \varphi_\text{high}(1-t).
\]

\subsection{Interplay between bridge and Wiener processes}

We will need below the well-known identity in law for the canonical bridge
\[
Y(t) \sim \mathcal{Y}(t): =(1-t) W\left({t\over 1-t}\right) .
\]
Using the change of time variable
\[
\tau = {t \over 1-t} \qquad \iff \qquad t = {\tau \over 1+\tau}
\]
and the scaling properties of the Wiener process,
we can replace the compounded process
\[
\mathcal{Y}(t(\tau))=\mathcal{Y}\left({\tau\over1+\tau}\right)
\]
by the more convenient process, which is identical in law and reads
\begin{equation}\label{mzthrywienerdef}
\mathcal{Y}(t(\tau)) \sim \mathcal{Z}(\tau) = {1\over1+\tau} W(\tau) .
\end{equation}
In turn, the following identity in law holds
\begin{equation}\label{yzequivalence}
Y(t) \sim \mathcal{Z}(\tau(t)) = \mathcal{Z}\left({t\over 1-t}\right) .
\end{equation}

\section{Joint pdf of the high value and its occurrence time}

\setcounter{equation}{0}
\setcounter{theorem}{0}
\setcounter{remark}{0}
\setcounter{lemma}{0}
\setcounter{definition}{0}
\setcounter{example}{0}
\setcounter{Assumption}{0}

\subsection{Reflection method}

Let us consider the function $f(\omega;\tau,h)$ such that
\begin{equation}\label{probsequaility}
\Pr\{W(\tau)\in(\omega,\omega+d\omega)\cap W(\tau')<h (1+\tau'): \tau'\in(0,\tau)\}= f(\omega;\tau,h)d\omega .
\end{equation}
This function $f(\omega;\tau,h)$ satisfies to the following diffusion equation
\begin{equation}\label{fomegaeq}
{\partial f \over \partial \tau} =  {1 \over 2} {\partial^2 f \over \partial \omega^2}
\end{equation}
with initial and absorbing boundary conditions
\begin{equation}\label{omegaconds}
f(\omega;\tau=0,h) = \delta(\omega) , \qquad f(\omega=h+h\tau;\tau,h) = 0 .
\end{equation}

We solve the initial-boundary problem \eqref{fomegaeq} with \eqref{omegaconds}
using the reflection method, which amounts to searching for
a solution of the form
\[
f(\omega;\tau,h) = {1 \over \sqrt{2\pi\tau}} \left[ \exp\left( - {\omega^2 \over 2\tau}\right) - A \exp\left( - {(\omega-2 h)^2 \over 2\tau}\right) \right] ,
\]
where the factor $A$ is defined from the absorbing boundary condition \eqref{omegaconds}, i.e.
\[
\exp\left( - {(h+h \tau)^2 \over 2\tau}\right) = A \exp\left( - {(h-h\tau)^2 \over 2\tau}\right) \quad \Rightarrow \quad A = e^{-2 h^2} .
\]
We thus obtain
\begin{equation}\label{fbridgemax}
f(\omega;\tau,h) = {1 \over \sqrt{2\pi\tau}} \left[ \exp\left( - {\omega^2 \over 2\tau}\right) - \exp\left( -2 h^2 - {(\omega-2 h)^2 \over 2\tau}\right) \right] .
\end{equation}

\subsection{Pdf of the maximal value of the canonical bridge}

In view of \eqref{probsequaility} and \eqref{mzthrywienerdef},
the joint pdf of $W(\tau)$ and  of the high value
\begin{equation}\label{htauforz}
\mathcal{H}(\tau) = \sup_{\tau'\in(0,\tau)} \mathcal{Z}(\tau')
\end{equation}
of the stochastic process $\mathcal{Z}(\tau')$ within the interval $\tau'\in(0,\tau)$ is equal to
\[
\mathcal{Q}(\omega,h;\tau) = {\partial f(\omega;\tau,h) \over \partial h} .
\]
Substituting in the above equation the expression \eqref{fbridgemax} yields
\begin{equation}\label{qomegahbridge}
\begin{array}{c} \displaystyle
\mathcal{Q}(\omega, h;\tau) = {1 \over \tau} \sqrt{{2 \over \pi \tau}} (2 h (1+\tau)-\omega) \exp\left[-2 h^2 - {(\omega-2 h)^2 \over 2\tau} \right] ,
\\[4mm] \displaystyle
\omega < h (1+\tau) , \qquad h>0 .
\end{array}
\end{equation}
In particular, the pdf of the high value $\mathcal{H}(\tau)$ \eqref{htauforz}
\[
\mathcal{Q}(h;\tau) = \int_{-\infty}^{h (1+\tau)} \mathcal{Q}(\omega,h,\tau) d\omega
\]
is equal to
\[
\mathcal{Q}(h;\tau) = \sqrt{{2 \over\pi\tau}} \exp\left(-{h^2 (1+\tau)^2 \over 2\tau}\right) +
2 h e^{-2 h^2} \text{erfc}\left({h (1-\tau) \over\sqrt{2\tau}}\right) .
\]
Using the identity in law \eqref{yzequivalence}, the pdf $Q_\text{high}(h;t)$ of the high value
\[
H(t) = \sup_{t'\in(0,t)} Y(t')
\]
is equal to
\begin{equation}\label{highbrpdf}
Q_\text{high}(h;t) = \mathcal{Q}\left(h;{t\over 1-t}\right) .
\end{equation}
In particular, the pdf's of the high values $\mathcal{H}$ \eqref{htauforz} and $H$ \eqref{highlowdef} are the same and equal to
\begin{equation}\label{pdfhigh}
\varphi_\text{high}(h) = \lim_{\tau\to\infty} \mathcal{Q}(h;\tau) = 4 h e^{-2 h^2} .
\end{equation}

\subsection{Pdf of the high value of the bridge and of its occurrence value}

In order to derive the joint pdf of the maximal value $H$ \eqref{highlowdef} and
of the occurrence time $t_\text{high}$ \eqref{instmaxmin}, we first
consider the related joint pdf of the high value $\mathcal{H}$ \eqref{htauforz} of
the auxiliary process $\mathcal{Z}(\tau)$ \eqref{mzthrywienerdef} and
of its occurrence time $\tau_\text{high} : ~ \mathcal{H} = \mathcal{Z}(\tau_\text{high})$.

The function $F(h,\tau)$ that defines the probability
\[
F(h,\tau)dh = \Pr\{\mathcal{H}\in(h,h+dh),\tau_\text{high}< \tau\} .
\]
is given by
\begin{equation}\label{fhtauint}
F(h,\tau) = \int_{-\infty}^{h(1+\tau)} \mathcal{Q}(\omega,h;\tau) P(\omega,h,\tau) d\omega ,
\end{equation}
where $\mathcal{Q}(\omega,h;\tau)$ is the joint pdf of $W(\tau)$ and $\mathcal{H}(\tau)$, given by equality \eqref{qomegahbridge}, while
\begin{equation}\label{peqprobomhtheta}
\begin{array}{c}\displaystyle
P(\omega,h,\tau) = \lim_{\theta\to\infty} P(\omega,h,\tau,\theta) ,
\\[3mm]\displaystyle
P(\omega,h,\tau,\theta) =\Pr\{W(\tau'|\tau,\omega)<h(1+\tau'):\tau'\in(\tau,\tau+\theta)\} .
\end{array}
\end{equation}
Here, $W(\tau'|\tau,\omega)$ is the conditioned Wiener process
that takes the value $\omega$ at $\tau'=\tau$.
Due to the identity in law \eqref{mzthrywienerdef},
$P(\omega,h,\tau)$ is equal to the probability that the following inequality holds
\[
\mathcal{Z}(\tau'|\tau,\omega) < h , \qquad \tau'\in(\tau,\infty) ,
\]
where $\mathcal{Z}(\tau'|\tau,\omega)$ is the conditioned stochastic process $\mathcal{Z}(\tau')$,
which is equal to $\omega/(1+\tau)$ at $\tau'=\tau$.

The probability $P(\omega,h,\tau,\theta)$ \eqref{peqprobomhtheta} is given by
\begin{equation}\label{promhtth}
P(\omega,h,\tau,\theta)= \int_{-\infty}^{h(1+\tau+\theta)} f(x;\omega,h,\tau,\theta) d x ,
\end{equation}
where the pdf $f(x,\omega,h,\tau,\theta)$ satisfies the initial-boundary value problem
\[
\begin{array}{c}\displaystyle
{\partial f \over \partial \theta} = {1 \over 2} {\partial^2 f \over \partial x^2} ,
\\[4mm] \displaystyle
f(x;\omega,h,\tau,\theta=0) = \delta(x-\omega) , \quad
f(x=h(1+\tau+\theta);\omega,h,\tau,\theta) = 0 .
\end{array}
\]
Its solution, obtained by the reflection method, is
\[
\begin{array}{c}\displaystyle
f(x;\omega,h,\tau,\theta) = {1 \over \sqrt{2 \pi\theta}} \bigg[ \exp\left(-{(x-\omega)^2 \over 2 \theta}\right) -
\\[4mm]\displaystyle
\exp\left(-2 h (h (1+\tau)-\omega) - {(x+\omega-2 h(1+\tau))^2 \over 2\theta}\right) \bigg] .
\end{array}
\]
Substituting this last expression into \eqref{promhtth} yields
\[
\begin{array}{c}
P(\omega,h,\tau,\theta)=
\\[4mm]\displaystyle
{1 \over 2} \left[ \text{erfc}\left( {\omega -h(1+\tau+\theta) \over \sqrt{2\theta}}\right) - e^{-2 h (h(1+\tau)-\omega)} \text{erfc}\left( {h(1+\tau-\theta)-\omega \over \sqrt{2\theta}}\right) \right] .
\end{array}
\]
In particular, in the limiting case $\theta\to\infty$, one has
\begin{equation}\label{prowhthinf}
P(\omega,h,\tau) = 1 - e^{-2 h (h(1+\tau) -\omega)} .
\end{equation}
Substituting $\mathcal{Q}(\omega,h;\tau)$ \eqref{qomegahbridge} and $P(\omega,h,\tau)$ \eqref{prowhthinf} into \eqref{fhtauint}, after integration, we obtain
\begin{equation}\label{fhtauexplicit}
F(h,\tau) = 2 h e^{-2 h^2} \text{erfc}\left({h (1-\tau) \over \sqrt{2\tau}}\right) .
\end{equation}

Consider now the probability
\[
\Phi_\text{high}(h,t) dh = \Pr\{H\in(h,h+dh),t_\text{high}<t\} .
\]
Due to the identity in law \eqref{yzequivalence}, $\Phi_\text{high}(h,t)$ is equal to
\begin{equation}\label{phihtcumulative}
\Phi_\text{high}(h,t) = F\left(h, {t \over 1-t}\right) = 2 h e^{-2 h^2} \text{erfc}\left({h (1-2 t) \over \sqrt{2 t (1-t)}}\right) .
\end{equation}
The integration over $h\in(0,\infty)$ gives the cumulative distribution function (cdf) of
the random occurrence times $t_\text{high}$ \eqref{instmaxmin}:
\[
\Phi_\text{high}(t) = \Pr\{t_\text{high}<t\} = \int_0^\infty \Phi(h,t) dh =  t , \qquad t\in (0,1) .
\]
This means that the occurrence time $t_\text{high}$ of the high value of the
canonical bridge is uniformly distributed. The above cdf satisfies the symmetry property \eqref{cdfsreverse}. The corresponding pdf $\varphi_\text{high}(t)=1$ satisfies obviously to symmetry property \eqref{pdftsym}.

The sought joint pdf of the high value $H$ of canonical bridge $Y(t)$ and
of its corresponding occurrence time $t_\text{high}$ is
\begin{equation}\label{phithrougboldphi}
\varphi_\text{high}(h,t) = {\partial \Phi_\text{high}(h,t) \over \partial t} .
\end{equation}
Substituting here $\Phi_\text{high}(h,t)$ \eqref{phihtcumulative} yields
\begin{equation}\label{phijointpdfht}
\varphi_\text{high}(h,t) = \sqrt{2\over\pi} {h^2\over\sqrt{t^3 (1-t)^3}} \exp\left( - {h^2\over 2t(1-t)}\right) .
\end{equation}

\section{Statistics of the high, low and occurrence time of the last
extremum of the canonical bridge}

\setcounter{equation}{0}
\setcounter{theorem}{0}
\setcounter{remark}{0}
\setcounter{lemma}{0}
\setcounter{definition}{0}
\setcounter{example}{0}
\setcounter{Assumption}{0}

\subsection{Statistical description of the joint pdf of the high, low and occurrence time of the last extremum}

The occurrence times of the first and last absolute
extremes \eqref{instmaxmin} of canonical bridge $Y(t)$ are formally defined as
\begin{equation}\label{maxinsthighlow}
t_\text{first} = \inf\{t_L,t_H\} , \qquad t_\text{last} = \sup\{t_L,t_H\} ~.
\end{equation}
The joint pdf of the high $H$ and low $L$ \eqref{highlowdef}
together with the cdf of the occurrence time $t_\text{last}$ is given by
\begin{equation}\label{helpdftcdf}
\Phi_\text{last}(h,\ell,t) dh d\ell = \Pr\{H\in(h,h+dh) \cap L\in(\ell,\ell+d\ell) \cap t_\text{last}<t\}~ .
\end{equation}
We derive the function $\Phi_\text{last}(h,\ell,t)$ by using a natural
generalization of the reasoning presented in Appendix~B that led
to the joint pdf $\Phi_\text{high}(h,t)$ \eqref{phihtcumulative} of
the high value $H$ and of the cdf of the occurrence time $t_\text{high}$.
Namely, we calculate first the probability
\begin{equation}\label{fhltaudef}
F(h,\ell, \tau) dh d\ell= \Pr\{\mathcal{H}\in(h,h+dh),\mathcal{L}\in(\ell,\ell+d\ell),\tau_\text{last}< \tau\} ,
\end{equation}
where
\[
\begin{array}{c}\displaystyle
\mathcal{H} =\sup_{\tau\in(0,\infty)} \mathcal{Z}(\tau) , \qquad \mathcal{L} =\inf_{\tau\in(0,\infty)} \mathcal{Z}(\tau) ,
\\[2mm]
\tau_\text{last} = \sup\{\tau_\text{low},\tau_\text{high}\} , \quad \tau_\text{low}: ~ \mathcal{L}=\mathcal{Z}(\tau_\text{low}) , \quad \tau_\text{high}: ~ \mathcal{H}=\mathcal{Z}(\tau_\text{high}) .
\end{array}
\]

Analogously to \eqref{fhtauint}, $F(h,\ell,\tau)$ is equal to
\begin{equation}\label{fhtauexplicit}
F(h,\ell,\tau) = \int_{\ell (1+\tau)}^{h(1+\tau)} \mathcal{Q}(\omega,h,\ell,\tau) P(\omega,h,\ell,\tau) d\omega ,
\end{equation}
where
\begin{equation}\label{pdfqhell}
\mathcal{Q}(\omega,h,\ell,\tau) = - {\partial^2 f(\omega;h,\ell,\tau) \over \partial h \partial \ell}
\end{equation}
and the pdf $f(\omega;h,\ell,\tau)$ satisfies the initial-boundary problem
\begin{equation}\label{difeqphi}
\begin{array}{c}\displaystyle
{\partial f \over \partial \tau} = {1 \over 2} {\partial^2 f \over \partial \omega^2} , \qquad
f(\omega;h,\ell,\tau=0) = \delta(\omega) ,
\\[2mm]\displaystyle
f(\omega=h(1+\tau);h,\ell,\tau)= 0 , \quad f(\omega=\ell(1+\tau);h,\ell,\tau)= 0 , \quad \tau>0 .
\end{array}
\end{equation}
Similarly to $P(\omega,h,\tau)$ \eqref{peqprobomhtheta}, the probability
$P(\omega,h,\ell,\tau)$ is given by
\[
\begin{array}{c}\displaystyle
P(\omega,h,\ell,\tau) = \lim_{\theta\to\infty} P(\omega,h,\ell, \tau, \theta) ,
\\[3mm]\displaystyle
P(\omega,h,\ell,\tau,\theta) =\Pr\{\ell(1+\tau')<B\tau'|\tau,\omega)<h(1+\tau'):\tau'\in(\tau,\tau+\theta)\} .
\end{array}
\]
Analogously to \eqref{promhtth}, the last probability $P(\omega,h,\ell,\tau,\theta)$ is equal to
\begin{equation}\label{pomegahellint}
P(\omega,h,\ell,\tau,\theta) =\int_{\ell(1+\tau+\theta)}^{h(1+\tau+\theta)} f(x;\omega,h,\ell,\tau,\theta) d x ,
\end{equation}
where $f(x;\omega,h,\ell,\tau,\theta)$ is the solution of the initial-boundary problem
\begin{equation}\label{difeqphiomega}
\begin{array}{c}\displaystyle
{\partial f \over \partial \theta} = {1 \over 2} {\partial^2 f \over \partial x^2} ,
\quad
f(x;\omega,h,\ell,\tau,\theta=0) = \delta(x-\omega) ,
\\[3mm]\displaystyle
f(x=h(1+\tau+\theta);\omega,h,\ell,\tau,\theta) = 0 , \quad
f(x=\ell(1+\tau+\theta);\omega,h,\ell,\tau,\theta) = 0 .
\end{array}
\end{equation}

Knowing the function $F(h,\ell,\tau)$ defined by equality \eqref{fhltaudef}, one can
find the sought function $\Phi_\text{last}(h,\ell,t)$ \eqref{helpdftcdf} using the following relation
\begin{equation}\label{lastfthroughf}
\Phi_\text{last}(h,\ell,t) = F\left(h,\ell,{t\over 1-t}\right) ,
\end{equation}
which is analogous to \eqref{phihtcumulative}. In turn, one can find
the joint pdf of the high $H$, low $L$ values \eqref{highlowdef} and
occurrence time of the last absolute extremum $t_\text{last}$ \eqref{maxinsthighlow} of
the canonical bridge $Y(t)$ using, analogously to \eqref{phithrougboldphi}, the relation
\begin{equation}\label{phihltpdfexpr}
\varphi_\text{last}(h,\ell,t) = {\partial \Phi_\text{last}(h,\ell,t) \over \partial t} .
\end{equation}

\subsection{Solutions of boundary-value problems \label{hyjywtbbr}}

Using the initial-boundary problem \eqref{difeqphi} with the reflection method, we obtain
\begin{equation}\label{phiomtauexpr}
\begin{array}{c} \displaystyle
f(\omega;h,\ell,\tau) =
\sum_{m=-\infty}^\infty \big[ e^{-2(h-\ell)^2 m^2 }
g(\omega +2 (h-\ell)m;\tau)-
\\[4mm] \displaystyle
e^{-2 ((h-\ell) m + h)^2} g(\omega-2 (h+ (h-\ell)m);\tau) \big] ,
\end{array}
\end{equation}
where
\[
g(\omega;\tau) = {1 \over \sqrt{2\pi\tau}} \exp\left( - {\omega^2 \over 2\tau}\right) .
\]
In turn, the solution of the initial-boundary problem \eqref{difeqphiomega} is given by
\begin{equation}\label{phiomtauexpromega}
\begin{array}{c}\displaystyle
f(x;\omega,h,\ell,\tau,\theta) = \sum_{m=-\infty}^\infty \bigg[ e^{-2 (h - \ell)^2 m^2 (1+\tau) + 2 \omega (h-\ell) m } \times
\\[4mm]\displaystyle
g(y-\omega+2 m (h-\ell) (1+\tau);\theta) -
\\[2mm]\displaystyle
e^{-2((h-\ell) m+h)^2 (1+\tau)+2 \omega ((h-\ell)m+h)}
g(y+\omega-2 ((h-\ell)m +h)(1+\tau) ;\theta) \bigg] .
\end{array}
\end{equation}

After substituting $f(\omega;h,\ell,\tau)$ \eqref{phiomtauexpr} into \eqref{pdfqhell}, we obtain
\begin{equation}\label{qomhelldistr}
\begin{array}{c} \displaystyle
\mathcal{Q}(\omega,h,\ell,\tau) = {4 \over \tau^2} \sum_{-\infty}^\infty m
\bigg[ m e^{-2 (h-\ell)^2 m^2} \times
\\[4mm] \displaystyle
[(\omega+2 m(h-\ell)(1+\tau))^2 - \tau(1+\tau)] g(\omega+2 m (h-\ell),\tau) -
\\[4mm] \displaystyle
(1+m) e^{-2(m (h-\ell)+h)^2} \times
\\[2mm] \displaystyle
[(\omega- 2 (m (h-\ell)+h) (1+\tau))^2 -\tau(1+\tau)] g(\omega-2 (m(h-\ell)+h), \tau)\bigg] .
\end{array}
\end{equation}
Substituting  $f(x;\omega,h,\ell,\tau,\theta)$ \eqref{phiomtauexpromega} into \eqref{pomegahellint},
and taking the limit $\theta\to\infty$, we obtain
\begin{equation}\label{probomheltau}
\begin{array}{c} \displaystyle
P(\omega,h,\ell,\tau) =
\\[1mm]\displaystyle
 \sum_{m=-\infty}^\infty
\left[e^{-2 (h - \ell)^2 (1+\tau) m^2 + 2 (h-\ell) m \omega }
- e^{-2 (h+(h - \ell) m)^2 (1+\tau) + 2(h+ (h-\ell) m) \omega} \right] .
\end{array}
\end{equation}

After substituting $\mathcal{Q}(\omega,h,\ell,\tau)$ \eqref{qomhelldistr} and $P(\omega,h,\ell,\tau)$ \eqref{probomheltau} into \eqref{fhtauexplicit}, we obtain the explicit expression for $F(h,\ell,\tau)$.
Substituting it into \eqref{lastfthroughf} and using relation \eqref{lastfthroughf},
we obtain the pdf of the high $H$, low $L$  values and occurrence time $t_\text{last}$
of the last extremum under the form
\begin{equation}\label{fhelltexpr}
\begin{array}{c}\displaystyle
\varphi_\text{last}(h,\ell,t) = \sum_{m=-\infty}^\infty \sum_{n=-\infty}^\infty \Big( m^2 \Big[
\\[4mm]\displaystyle
g(h,t,2 (h-\ell) m,2 (h-\ell)n) - g(\ell,t,2 (h-\ell)m,2 (h-\ell)n) -
\\[4mm]\displaystyle
g(h,t,2 (h-\ell) m, 2 (h+(h-\ell)n)) +
g(\ell,t,2 (h-\ell)m, 2(h+(h-\ell)n))\Big] -
\\[3mm]\displaystyle
m (m+1) \Big[g(h,t,-2 (h+(h-\ell) m),2 (h-\ell)n) -
\\[3mm]\displaystyle
g(\ell,t,-2 (h+(h-\ell) m),2 (h-\ell)n) -
\\[3mm]\displaystyle
g(h,t,-2 (h+(h-\ell) m),2 (h+(h-\ell)n)) +
\\[3mm]\displaystyle
g(\ell,t,-2 (h+(h-\ell) m),2 (h+(h-\ell)n))\Big]\Big) ,
\\[3mm]\displaystyle
g(y,t,a,c) = - \sqrt{{2\over\pi(1-t)^3 t^7}} \exp\left(- {(a+y)^2-(a+c) (a-c+2 y) t\over 2 t (1-t)}\right) \times
\\[5mm] \displaystyle
\left[(a+y)^3 -(a+y)(3+(a+y)(a-c+2y)) t + (3a-c+4y) t^2 \right] .
\end{array}
\end{equation}

\subsection{Function \textnormal{$\alpha_\text{last}(\theta,t;\lambda)$} \label{jruki}}

Some of the most efficient estimators
introduced in this paper are defined through the function
\begin{equation}\label{gmudef}
\alpha_\text{last}(\theta,t;\lambda) =  \int_0^\infty r^{\lambda+1}\varphi_\text{last}(r \cos\theta,r\sin\theta,t) dr ,
\end{equation}
which is  analogous to \eqref{alpparkdef},
The calculation of the integral \eqref{gmudef} yields
\begin{equation}\label{alphlasttexpr}
\begin{array}{c}\displaystyle
\alpha_\text{last}(\theta,t;\lambda) = - {1\over\sqrt{8\pi(1-t)^3 t^7 \delta^{5+\lambda}}}~ \Gamma\left({3+\lambda\over 2}\right) \sum_{m,n=-\infty}^\infty \Big( m^2 \times
\\[5mm]\displaystyle
\Big[\beta(co,t,2 sc\cdot m,2 sc\cdot n;\lambda) -
\beta(si,t,2 sc\cdot m,2 sc\cdot n;\lambda) -
\\[2mm]\displaystyle
\beta(co,t,2 sc\cdot m, 2 (co+sc\cdot n);\lambda) +
\\[2mm]\displaystyle
\beta(si,t,2 sc\cdot m, 2(co+sc\cdot n);\lambda)\Big] -
\\[2mm]\displaystyle
m (m+1) \Big[\beta(co,t,-2 (co+sc\cdot m),2 sc\cdot n;\lambda) -
\\[2mm]\displaystyle
\beta(si,t,-2 (si+sc\cdot m),2 sc\cdot n;\lambda) -
\\[2mm]\displaystyle
\beta(co,t,-2 (co+sc\cdot m),2 (co+sc\cdot n);\lambda) +
\\[2mm]\displaystyle
\beta(si,t,-2 (co+sc\cdot m),2 (co+sc\cdot n);\lambda)\Big]\Big) ,
\end{array}
\end{equation}
where
\[
\begin{array}{c}
\displaystyle
co = \cos\theta , \qquad si = \sin\theta , \qquad sc = co-si ,
\\[2mm]
\displaystyle
\beta(y,t,a,c;\lambda) =
\Big[(a+y)^2[a+y-(a-c+2y)t] (3+\lambda) +
\\[2mm]\displaystyle
\delta t[(6a-2c+8y)t- 6(a+y)] \Big] ,
\\[4mm]\displaystyle
\delta=\delta(y,t,a,c) = {(a+y)^2-(a+c) (a-c+2 y) t\over 2 t (1-t)} .
\end{array}
\]

\subsection{Function \textnormal{$\alpha(\theta,\upsilon,t;\lambda)$} \label{hruyjkikik}}

Consider the function
\begin{equation}\label{athuptlaga}
\alpha(\theta,\upsilon,t;\lambda) = \int_0^\infty r^{\lambda+2} \varphi_\text{last}(r\cos\upsilon\cos\theta,r\cos\upsilon\sin\theta,r\sin\upsilon,t; \gamma=0) dr ,
\end{equation}
that enters into the definition of the
canonical estimator \eqref{bthtmin} in the case of zero drift $\gamma=0$.
Using expression \eqref{philasthelxtgam} for the pdf $\varphi_\text{last}(h,\ell,x,t; \gamma)$,
we obtain after calculations the following expression
\begin{equation}\label{alphlasttexprprim}
\begin{array}{c}\displaystyle
\alpha(\theta,\upsilon,t;\lambda) = {\Gamma\left({4+\lambda\over2}\right)\over 4 \pi \sqrt{\mathstrut t^7 (1-t)^3}}\sum_{m,n=-\infty}^\infty \Big( m^2 \times
\\[2mm]\displaystyle
\Big[\beta'(x,co,t,sc\cdot m, sc\cdot n;\lambda) -
\beta'(x,si,t, sc\cdot m,sc\cdot n;\lambda) -
\\[2mm]\displaystyle
\beta'(x,co,t,sc\cdot m, cc+sc\cdot n;\lambda) +
\beta'(x,si,t,sc\cdot m, cc+sc\cdot n;\lambda)\Big]
-
\\[3mm]\displaystyle
m (m+1) \Big[\beta'(x,co,t,-cc-sc\cdot m,sc\cdot n;\lambda) -
\\[2mm]\displaystyle
\beta'(x,si,t,-cc-sc\cdot m,sc\cdot n;\lambda) -
\\[3mm]\displaystyle
\beta'(x,co,t,-cc-sc\cdot m,cc+sc\cdot n;\lambda) +
\\[2mm]\displaystyle
\beta'(x,si,t,-cc-sc\cdot m,cc+sc\cdot n;\lambda)\Big]\Big) .
\end{array}
\end{equation}
which is analogous to \eqref{alphlasttexpr}.
Here, we have set
\[
\begin{array}{c}
x =\sin\upsilon, \qquad co = \cos\theta \cos\upsilon, \qquad si =\sin\theta \cos\upsilon ,
\\[1mm]
cc= 2 \cos\theta \cos\upsilon , \qquad sc = 2 (\cos\theta-\sin\theta) \cos\upsilon ,
\\[3mm]\displaystyle
\beta'(x,y,t,a,c,\lambda) =
\big[r(4+\lambda) (a+y-(a-c+2 y) t)+
\\[2mm]\displaystyle
\delta t ((6 a-2 c+8 y) t -6 (a+y))\big] \delta^{-(6+\lambda)/2} ,
\\[3mm]\displaystyle
\delta = {r-(a+c) (a-c+2 y) t \over 2 t (1-t)} + {x^2\over2}, \qquad r = (a+y)^2 .
\end{array}
\]

\clearpage

\section*{References}

A\"it-Sahalia, Y.,  P.A. Mykland, and L.~Zhang (2005).
How often to sample a continuous-time process in the presence of market microstructure noise. \emph{Review of Financial Studies} 18, 351-416.

Andersen, T.~G., T. Bollershev, F. X. Diebolt and P. Labys (2003). Modeling and Forecasting Realized Volatility. {\em Econometrica} 71, 529-626.

Garman, M. and M. J. Klass (1980).
On the Estimation of Security Price Volatilities From Historical Data.
{\em Journal of Business} 53, 67-78.

Jeanblanc J. \& M. Yor, and M. Chesney (2009).
\emph{Mathematical Methods for Financial Markets.} Springer.

Parkinson, M. (1980).
The Extreme Value Method for Estimating the Variance of the Rate of Return. {\em Journal of Business} 53, 61-65.

Saichev, A., D. Sornette, V. Filimonov F. Corsi (2009).
Homogeneous Volatility Bridge Estimators.
{\em ETH Zurich working paper}, \url{http://ssrn.com/abstract=1523225}.

Saichev A., Y. Malevergne, D. Sornette (2010) \emph{Theory of Zipf's Law and Beyond} (Lecture Notes in Economics and Mathematical Systems), Springer.

Zhang, L., Mykland, P.A. and A•t-Sahalia, Y. (2005).
A tale of two time scales: determining integrated volatility with noisy high-frequency data.
\emph{Journal of the American Statistical Association} 100, 1394-1411.

\clearpage

\begin{quote}
\centerline{
\includegraphics[width=14cm]{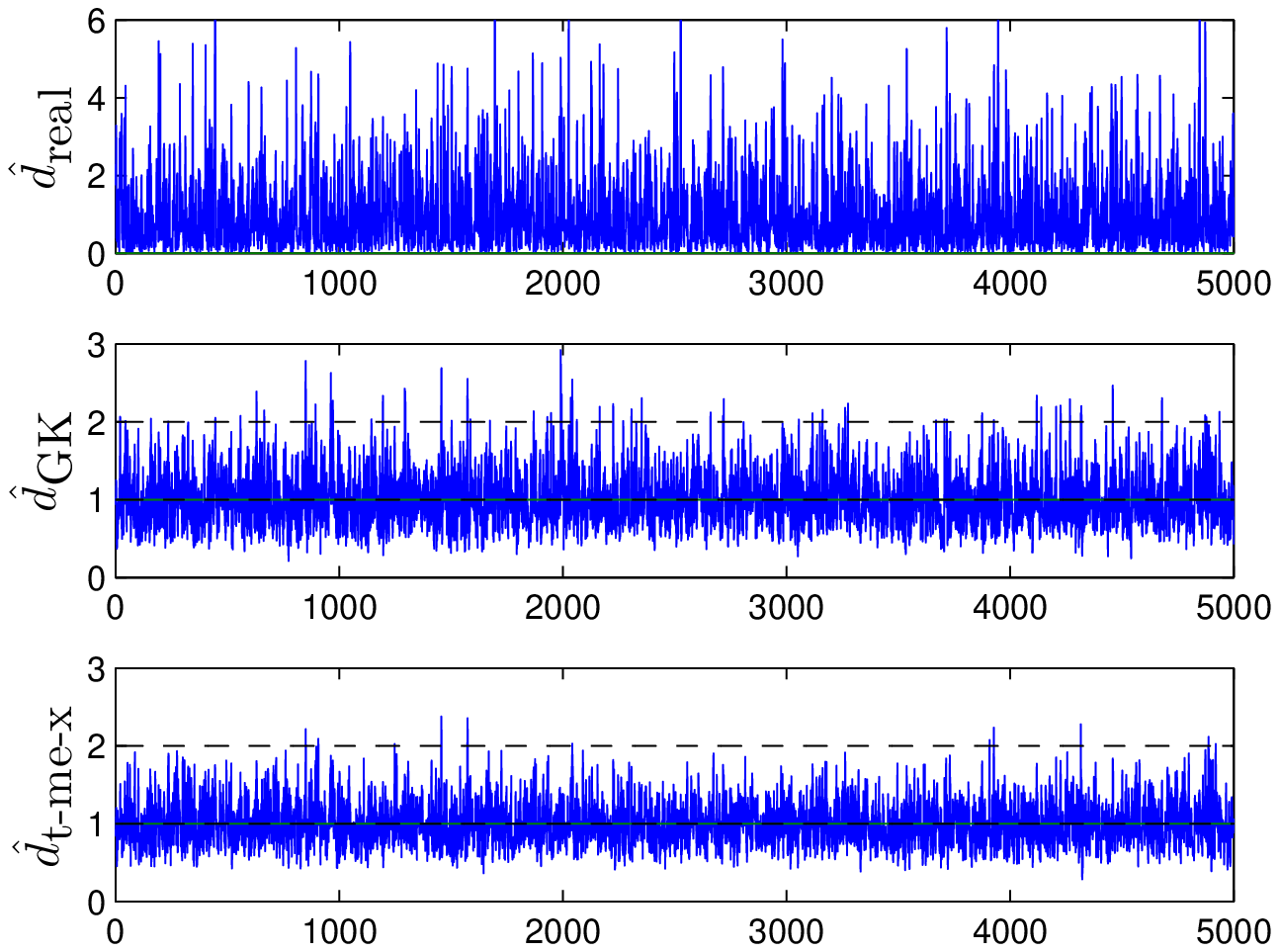}}
{\bf Fig.~1:} \small{5000 realizations  of the open-close estimator
 \eqref{realspotnumsimest}, of the G\&K estimator  \eqref{gkspotnumsimest}, and
 of the most efficient estimator $\hat{d}_\text{t-me-x}$ \eqref{bthtmin}.}
\end{quote}

\clearpage

\begin{quote}
\centerline{
\includegraphics[width=14cm]{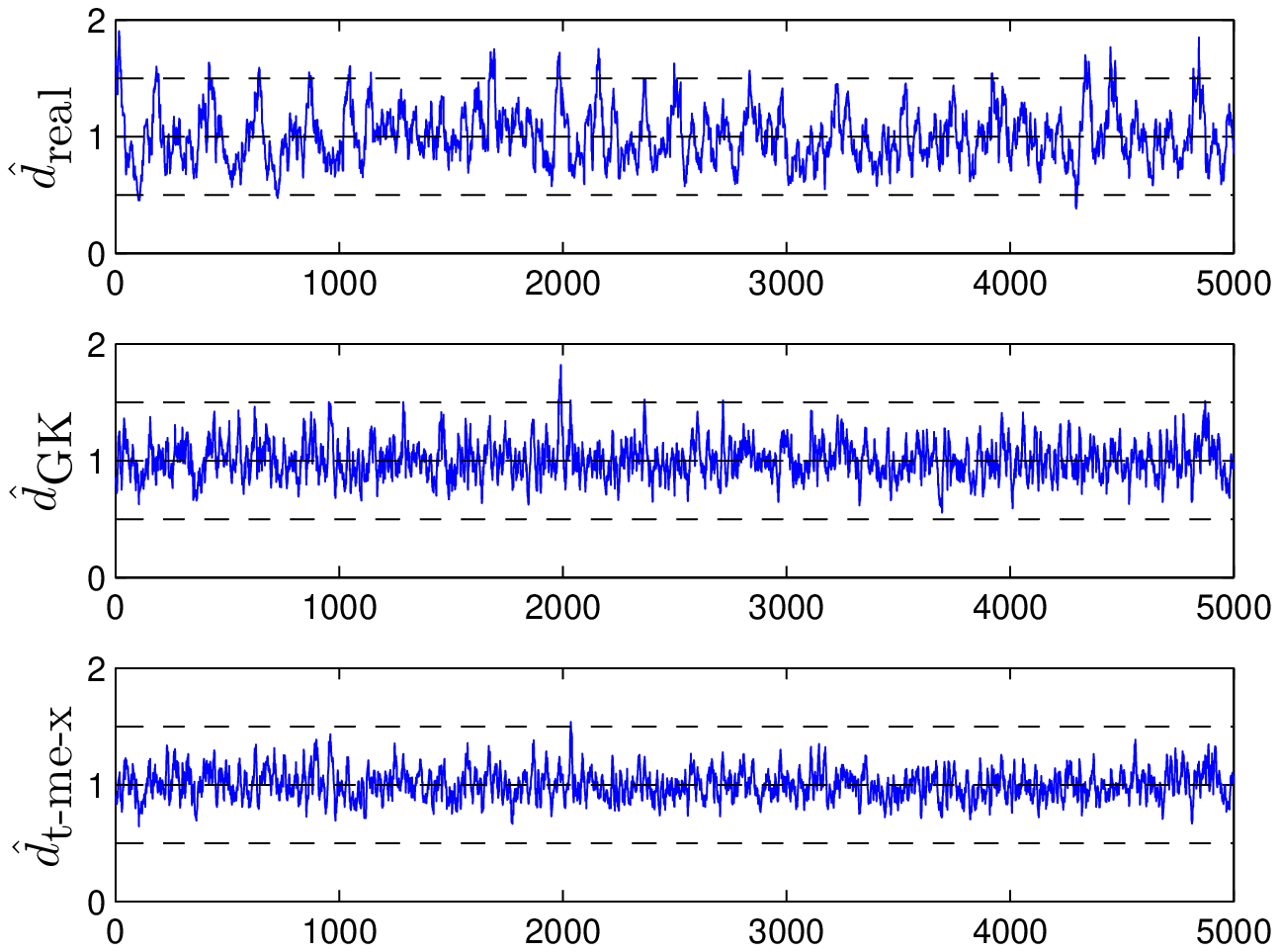}}
{\bf Fig.~2:} \small{Moving averages of the open-close (top panel), G\&K
(middle panel) and time-OHLC \eqref{bthtmin} (lower panel) estimators
over respective windows sizes of 30 samples for the top panel and 10 samples
for the two other panels. As explained in the text, this
moving average mimicks
the normalized estimators of the integrated variance
in the case where all instantaneous variances are the same ($\sigma_i^2=\sigma^2=\text{const}$).
}
\end{quote}

\clearpage

\begin{quote}
\centerline{
\includegraphics[width=13cm]{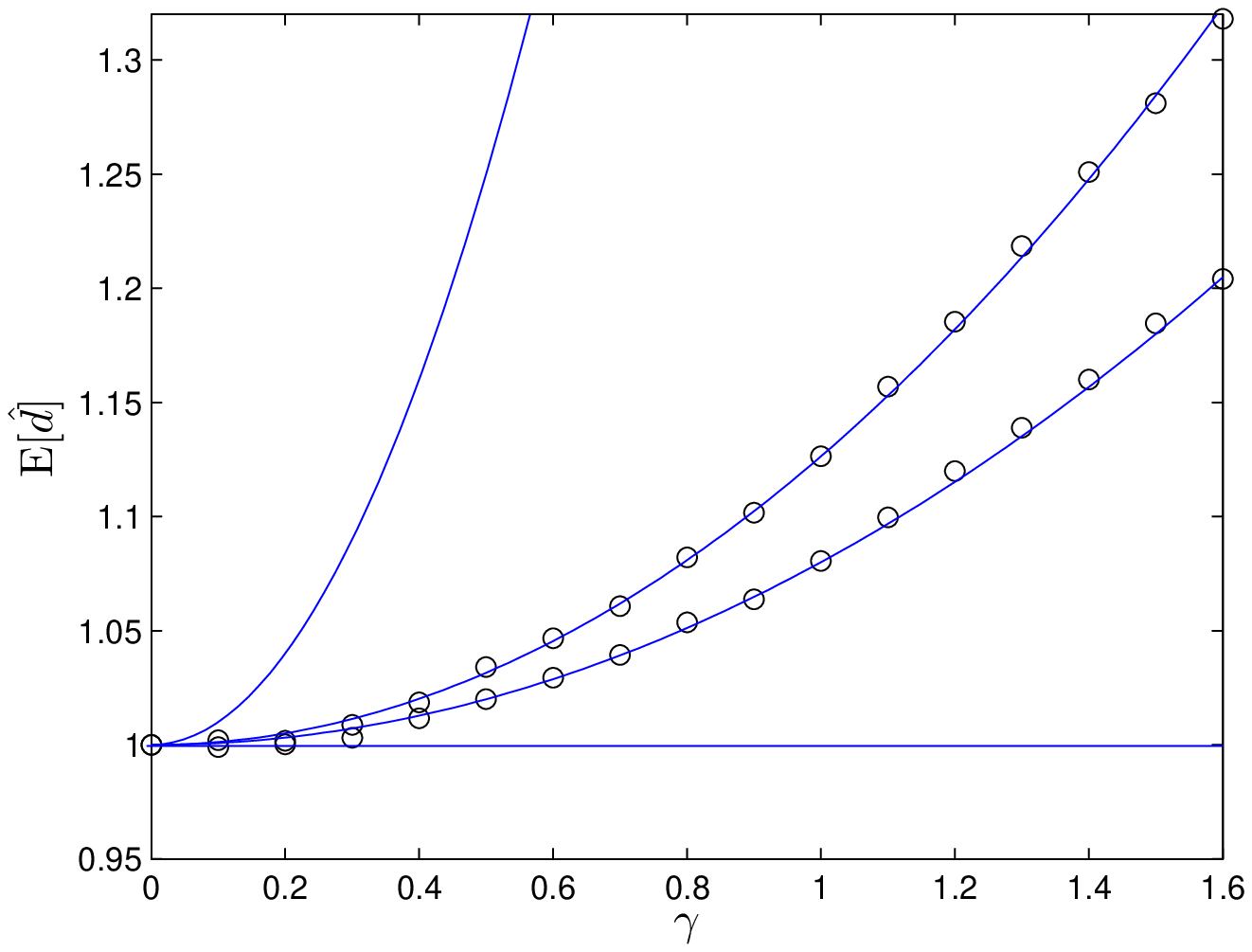}}
{\bf Fig.~3:} \small{Top to bottom, $\gamma$-dependence
of the expected values of the open-close $\hat{d}_\text{real}$ \eqref{realspotnumsimest}, G\&K $\hat{d}_\text{GK}$ \eqref{gkspotnumsimest} and most efficient $\hat{d}_\text{t-me-x}$ \eqref{bthtmin} canonical estimators. The horizontal line is the expected value of the canonical estimators $\hat{d}_\text{bPark}$ and $\hat{d}_\text{t-me}$}
\end{quote}

\clearpage

\begin{quote}
\centerline{
\includegraphics[width=13cm]{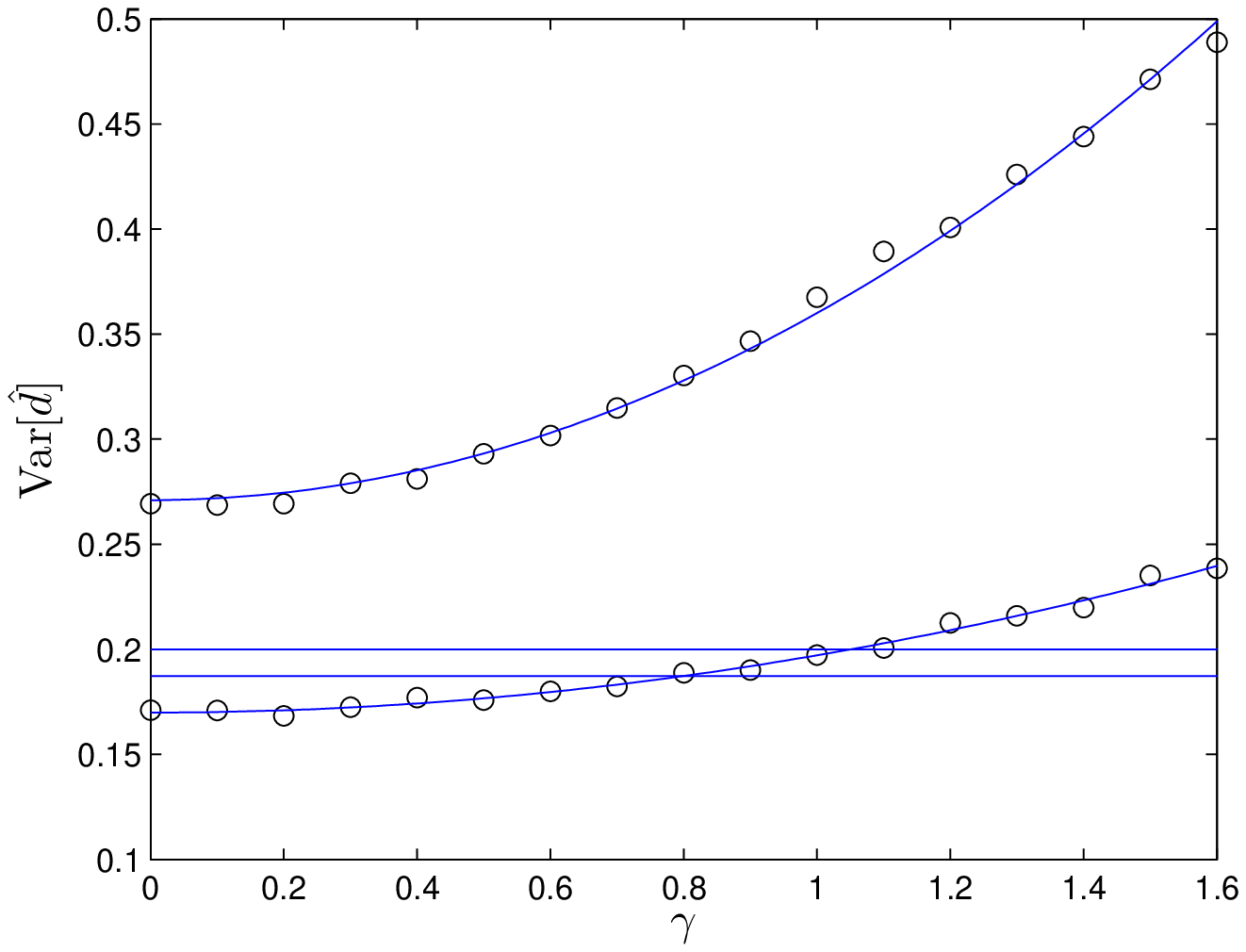}}
{\bf Fig.~4:} \small{$\gamma$-dependence of the statistical average of the
variances of the canonical estimators $\hat{d}_\text{GK}$ (upper open circles) and $\hat{d}_\text{t-me-x}$
(lower open circles).
The two horizontal lines are the variances \eqref{parktmevars} of
the canonical estimators $\hat{d}_\text{bPark}$ (top) and $\hat{d}_\text{t-me}$ (bottom), respectively.}
\end{quote}

\end{document}